\def\etal{{\it et al.\/\ }}

\def\hii{\ion{H}{2}}
\def\sun{$_\odot$}

\def\arcsec{^{\prime\prime}}
\documentstyle[11pt,aaspp4,epsf]{article}
\lefthead{GARNETT ET AL.}
\righthead{CARBON IN SPIRAL GALAXIES}
\accepted{30 September 1998}
\slugcomment{To appear in Ap. J. }
\begin{document}
\title{CARBON IN SPIRAL GALAXIES FROM {\it HUBBLE SPACE TELESCOPE} 
SPECTROSCOPY\altaffilmark{1}}
\author{D. R. Garnett\altaffilmark{2,8}, G. A. Shields\altaffilmark{3},
M. Peimbert\altaffilmark{4}, S. Torres-Peimbert\altaffilmark{4}, 
E. D. Skillman\altaffilmark{2}, R. J. Dufour\altaffilmark{5}, E.
Terlevich\altaffilmark{6}, and R. J. Terlevich\altaffilmark{7}}
\altaffiltext{1}{Based on observations with the NASA/ESA Hubble Space
Telescope obtained at Space Telescope Science Institute, which is operated
by the Association of Universities for Research in Astronomy, under NASA
contract NAS5-26555.}
\altaffiltext{2}{Astronomy Department, University of Minnesota, Minneapolis 
MN 55455} 
\altaffiltext{3}{Astronomy Department, University of Texas, Austin, TX 78712}
\altaffiltext{4}{Instituto de Astronom\'\i a, UNAM, Apdo Postal 70-264, 
M\'exico 04510 D.F., M\'exico}
\altaffiltext{5}{Department of Space Physics and Astronomy, Rice University, 
Houston, TX 77251-1892}
\altaffiltext{6}{Institute of Astronomy, Madingley Road, Cambridge 
CB3 0HA, United Kingdom}
\altaffiltext{7}{Royal Greenwich Observatory, Madingley Road, Cambridge 
CB3 0EZ, United Kingdom}
\altaffiltext{8}{Current address: Steward Observatory and Astronomy Department, 
University of Arizona, Tucson, AZ 85721; e-mail: garnett@oldstyle.spa.umn.edu}
\begin{abstract}
We present measurements of the gas-phase C/O abundance ratio in six \hii\ 
regions in the spiral galaxies M101 and NGC 2403, based on ultraviolet 
spectroscopy using the Faint Object Spectrograph on the {\it Hubble Space 
Telescope}. The C/O ratios increase systematically with O/H in both galaxies, 
from log C/O $\approx$ $-$0.8 at log O/H = $-$4.0 to log C/O $\approx$ 
$-$0.1 at log O/H = $-$3.4. C/N shows no correlation with O/H. The rate 
of increase of C/O is somewhat uncertain because of uncertainty as to the 
appropriate UV reddening law, and uncertainty in the metallicity dependence
on grain depletions. However, the trend of increasing C/O with O/H is 
clear, confirming and extending the trend in C/O indicated previously from 
observations of irregular galaxies. Our data indicate that the radial 
gradients in C/H across spiral galaxies are steeper than the gradients 
in O/H. Comparing the data to chemical evolution models for spiral galaxies 
shows that models in which the massive star yields do not vary with 
metallicity predict radial C/O gradients that are much flatter than the 
observed gradients. The most likely hypothesis at present is that stellar 
winds in massive stars have an important effect on the yields and thus on 
the evolution of carbon and oxygen abundances. C/O and N/O abundance ratios 
in the outer disks of spirals determined to date are very similar to those 
in dwarf irregular galaxies. This implies that the outer disks of spirals 
have average stellar population ages much younger than the inner disks.

\end{abstract}

\section{Introduction}

The abundance of carbon in galaxies and its evolution relative to oxygen
provides fundamental information for understanding a variety of problems 
in stellar evolution, galaxy evolution, and the interstellar medium (ISM). 
A large fraction of carbon is produced in intermediate-mass stars (carbon 
star and planetary nebula progenitors), while oxygen is synthesized 
almost entirely in stars above 10 M$_{\sun}$ (Maeder 1992\markcite{m1}; 
Woosley \& Weaver 1995, hereafter WW95\markcite{w1}; Renzini \& Voli 
1981\markcite{r1}). The carbon abundance thus traces element enrichment over 
much longer timescales than oxygen, and so the C/O ratio is potentially 
useful as an indicator of the time since the bulk of star formation has
occurred in a galaxy (e.g., Pagel \& Edmunds 1978\markcite{pe78}; Garnett 
et al. 1997a\markcite{g5}; hereafter G97a). Indeed, chemical evolution 
models for spiral galaxies predict very different behaviors for the 
evolution of C/O ratios (see discussion in Section 4 below). Additionally, 
cooling in dark clouds can be dominated by emission from carbon ions and 
carbon-bearing molecules (Tielens \& Hollenbach 1985\markcite{t1}; Shull 
\& Woods 1985\markcite{s1}), so the carbon abundance is highly relevant 
for the thermal balance in clouds. The variation of C/H and C/O in the 
ISM is important for modeling the formation of CO in molecular clouds 
and the possible effects of abundance variations on the I(CO)/N(H$_2$) 
relation. Evidence from the Galaxy and the Magellanic Clouds indicates 
that N(H)/E(B$-$V) varies directly with C/H (Mathis 1990\markcite{m2}), 
so the carbon abundance in the ISM is a potential predictor of the dust 
to gas ratio in galaxies.

Little data has been available for carbon abundances in the ISM of other 
galaxies, particularly spiral galaxies, because relatively strong emission 
lines from carbon in the ionized gas, particularly emission from C$^{+2}$, 
can be observed only in the ultraviolet spectral region. A small number 
of $IUE$ observations of \hii\ regions in spirals have been made (Dufour, 
Schiffer, \& Shields 1984\markcite{d3}), but those data were often affected 
by low signal/noise, uncertainty in the reddening correction, and the 
uncertainty in matching the $IUE$ emission line strengths to ground-based 
spectra. $HST$ spectroscopy with the Faint Object Spectrograph (FOS) has 
improved this situation through the combination of larger dynamic range, 
improved detector noise characteristics, better spectral resolution, and 
the ability to measure both UV and optical emission lines with the same 
instrument and aperture (Dufour et al. 1993\markcite{d4}).

We have been carrying out a program of UV/optical spectroscopy of \hii\ 
regions in a variety of galaxies with $HST$ to derive reliable information 
on carbon abundances over a wide range of O/H. Recently completed studies 
of carbon abundances in metal-poor irregular galaxies have appeared in 
Garnett et al. (1995\markcite{g3}; hereafter G95), G97a, Kobulnicky et al. 
(1997)\markcite{k7}, and Kobulnicky \& 
Skillman (1998)\markcite{ks98}. In this paper, 
we present the results of FOS spectroscopy of \hii\ regions in the two 
spiral galaxies NGC 2403 and M101, from which we derived C/H and C/O 
abundance ratios. Sections 2 and 3 describe the observations and analysis 
of the spectra. Section 4 discusses the resulting abundance patterns, 
comparing the spiral galaxy data with our published data for irregular 
galaxies; we also compare the spiral galaxy data with published chemical 
evolution models for spiral disks, and discuss the results in the context 
of disk evolution and stellar nucleosynthesis.

\section{Observations and Reductions}

\subsection{FOS Spectroscopy}

We observed three \hii\ regions in M101 and three in NGC 2403 with the FOS
on {\it HST} during Cycle 5, using the red Digicon. For all six regions 
we scanned the full wavelength region from 1600 \AA\ to 6900 \AA\ with the 
four medium resolution gratings G190H, G270H, G400H, and G570H, observed 
through the 0$\arcsec$.86 square (C-1) upper aperture. The beam-switch 
option was disabled for these observations since sky background is not 
a significant source of error for these observations (except for possible 
skyglow contamination at [O~II] 2470 \AA). A journal of the FOS observations 
is provided in Table 1. 

All six \hii\ region targets were positioned in the FOS aperture by blind 
offset from nearby Guide Star Catalog stars with well-measured positions. 
The offsets were determined from astrometry of broad-band and narrow-band 
H$\alpha$ CCD images. We attempted to maximize the signal/noise in the 
UV emission lines by locating regions with the highest H$\alpha$ emission 
equivalent widths in the images. However, the acquisition and slew introduces 
some uncertainty in the final pointing, and the resulting s/n in the 
UV lines varied considerably. We were able to detect both the C~II] 2325 
\AA\ and C~III] 1907,1909 \AA\ emission features in four objects, while in 
one case we detected only C~II], and in another we detected only C~III].

The spectra were initially processed using the standard FOS pipeline 
reduction routines. Inspection of the pipeline products showed that
the flat-fielding was suspect; in particular, a well-known flat field
artifact at 1950 \AA\ was prominent in absorption in the G190H spectra.
In fact, the pipeline products had been processed with flats for the
C-1 aperture taken in 1992 (pre-COSTAR); since that time the 1950 \AA\
feature has disappeared. We subsequently reprocessed the spectra using 
flat-fields taken through the C-1 aperture in July 1996, which improved 
the spectra considerably. 

We measured fluxes for the emission lines by direct integration of the 
emission line profiles, with Gaussian profile fits used for comparison 
and to measure blended lines. Raw measured line fluxes relative to the 
H$\beta$ line are listed in Table 2. The 1$\sigma$ uncertainties in 
the raw line fluxes were determined by adding in quadrature the error 
contributions from the following sources: the statistical noise in the 
lines plus local continuum, determined from the raw counts, and the 
uncertainty in the photometric calibration of the FOS, approximately 
3\% (Bohlin 1995\markcite{b2}). Uncertainty due to gross variable 
features in the flat-field correction is not included. Statistical 
noise dominates the uncertainty in the C~III] and C~II] line 
fluxes. The upper limits listed in Table 2 represent 2$\sigma$ 
determinations based on statistical fluctuations in the nearby 
continuum.

\subsection{Corrections for Interstellar Reddening}

Because most of the \hii\ regions observed here are relatively metal-rich
objects in spiral galaxies, and because we must compare the fluxes of UV 
emission lines relative to the optical spectrum, interstellar reddening 
and obscuration are important issues. Interstellar reddening functions 
toward \hii\ complexes in other spiral galaxies are in fact not well 
understood. Metallicity, the intensity of star formation, the distribution
of reflecting and absorbing grains, and the size of the region over which 
the reddening is measured may affect the shape of the obscuration curve. 
The UV reddening law in the Galaxy shows significant variations with 
line of sight (see Cardelli, Clayton, \& Mathis 1989\markcite{ccm89}), 
with dense clouds associated with star formation generally showing greyer 
UV reddening curves. Bianchi \etal (1996)\markcite{b1} derived an average 
extinction law for stars in M31; they 
found an average extinction law similar to that in the Galaxy, except 
possibly for a weaker 2175 \AA\ bump. Rosa \& Benvenuti (1994)\markcite{r3} 
found a similar result based on FOS spectroscopy of 
OB associations in M101. The 
uncertainties in both of these results are still relatively large, however. 
Meanwhile, Calzetti, Kinney, \& Storchi-Bergmann (1994)\markcite{c1} 
derived a mean obscuration curve for starburst regions which 
is quite gray relative 
to the Galactic obscuration curve, with a very weak or absent 2175 
\AA\ feature; furthermore, they demonstrated that the obscuration 
toward the stars tends to be quite different from that toward the 
ionized gas, a likely consequence of clumping of the interstellar gas 
and different spatial distributions for the OB stars and the ionized gas.

We attempted to determine the characteristics of the reddening curves
for our NGC 2403 and M101 regions in several ways. We infer immediately
that the reddening law is something like a Galactic law, because the FOS 
spectra showing strong stellar continuum also show a prominent 2175 \AA\ 
bump. Therefore, we eliminate the SMC (Pr\' evot \etal 1984\markcite{p3}) 
and Calzetti \etal (1994)\markcite{c1} obscuration functions from further 
consideration.

Cardelli \etal (1989)\markcite{ccm89} demonstrated that the obscuration 
function in the UV could be characterized 
as a smoothly varying function of $R_V$, the ratio of general 
extinction to selective extinction. Lines of sight with large $R_V$ 
($\approx$ 5) have relatively grey UV reddening, while those with lower 
$R_V$ ($\approx$ 3) have relatively steep UV reddening curves. Most lines 
of sight have $R_V$ between these two values. Therefore, we consider these 
two values as limits on the range of possible reddening functions for 
our target \hii\ regions. 

The most straightforward approach would be to compare optical/UV lines
which have ratios fixed by atomic physics. He~I recombination lines
are a potentially very useful tool to derive the shape of the reddening
curve between 2500 \AA\ and 6700 \AA. We detect He~I 3187 \AA\ and 
He~I 2946 \AA\ in some of our spectra, which, when combined with 
measurements of optical He I lines, can potentially distinguish between 
flat and steep UV reddening curves. Unfortunately, the signal/noise 
in our He~I line measurements was insufficient to distinguish between 
$R_V$ = 3 and $R_V$ = 5. Improved measurements of UV He emission lines 
would be invaluable to constrain the reddening law in extragalactic 
\hii\ regions.

A second approach is to use the slope of the UV continuum to constrain 
the shape of the reddening curve. Calzetti \etal (1994) \markcite{c1} 
demonstrated that the slope $\beta$ (where $F_{\lambda} \propto 
\lambda^{\beta}$) of the UV continuum of a starburst is strongly 
correlated with the amount of interstellar obscuration. Starbursts 
that have little obscuration have UV slopes similar to theoretical 
predictions for young OB stellar populations ($\beta$ $\approx$ $-$2.5; 
Leitherer \& Heckman 1995\markcite{l1}). Four of our six 
targets show stellar 
continuum (NGC 5461 and NGC 5471 do not); the four have continuum 
slopes $-$2.0 $<$ $\beta$ $<$ $-$2.25, measured over the range 
1600-2700 \AA. Assuming that the star clusters are very young OB 
associations with little contamination from the old field star 
population, these slopes are consistent with modest reddening. We 
attempted to determine if we could distinguish between $R_V$ = 3.1 
and $R_V$ = 5 from the dereddened slope of UV continuum. We applied 
$R_V$ = 3.1 and $R_V$ = 5 reddening functions to the spectra so that 
the 2175 \AA\ feature disappeared. (This implied a higher $A_V$ for 
the $R_V$ = 5 case.) In all cases, the measured dereddened continuum 
slopes were not significantly different for the two reddening laws, 
and were similar to the theoretical slope. We were thus unable to 
constrain the reddening curves. Foreground Galactic reddening has a 
minor effect, contributing $A_V$ $<$ 0.06 magnitude for both galaxies, 
as determined from the Burstein \& Heiles (1982)\markcite{bh82} maps. 
Observations shortward of 1600 
\AA\ are likely to provide better constraints than our near-UV spectra.

Given the uncertainty in the reddening curves, we have chosen to present 
two sets of abundance results, based on both the $R_V$ = 3.1 
and $R_V$ = 5 obscuration laws. The two sets of results should represent 
the maximum range of possible values for these spiral galaxy 
environments. As we shall see, the largest difference we see 
(for NGC 5461) is a factor of two in the derived C/O abundance ratio. 
While we find that the uncertainty in reddening law will not 
affect our conclusions regarding the general trend of C/O 
in spirals, we will discuss the effects on our analysis when significant.

The interstellar reddening corrections for the ionized gas were estimated 
from H$\alpha$/H$\beta$ ratios measured from the G570H spectra. 
In the four cases where significant stellar continuum is 
present, the Balmer line ratios were corrected for underlying absorption 
of about 1.5 \AA\ equivalent width; this value was chosen 
to give the lowest dispersion in derived A(H$\beta$) for the 
various line ratios. The observed line fluxes were corrected for 
reddening using the parameterization of Cardelli \etal (1989) \markcite{ccm89} 
for $R_V$ = 3.1 and $R_V$ = 5. The reddening-corrected line 
fluxes for both reddening 
functions are listed in Table 2; the uncertainties listed for 
these fluxes include an additional error term due to the uncertainty in 
$A(\lambda)$, as determined from the errors in the Balmer line 
ratios but not including the uncertainty in the shape of the 
reddening curve. 

\section{Abundance Analysis}

\subsection{Physical Conditions}

It is necessary to have measurements of the electron density 
$n_e$ and electron temperature $T_e$ in each \hii\ region in 
order to compute ion abundances directly. For our FOS spectra, it 
was not possible in general to make such measurements, since it 
was necessary to devote most of our observing time allocation to 
long UV exposures in order to detect the faint carbon lines. 
Exposure times for the optical grating settings were thus 
short, yielding only modest signal/noise for emission lines weaker 
than 10\% of the H$\beta$ line. Therefore, we adopted electron temperatures 
and densities 
from published spectroscopy. The measured values for $n_e$ and 
$T_e$ are listed in Table 3. High-quality spatially resolved spectroscopy 
of giant \hii\ regions (D\'\i az \etal 1987\markcite{d2}; Kinkel 
1993\markcite{k5}; Kobulnicky \& Skillman 1996\markcite{k6}) show 
that $T_e$ varies little across such 
regions over size scales comparable to the FOS aperture and 
larger; $T_e$ variations over small size scales can not be 
ruled out, however.

For the NGC 2403 objects, we generally used the temperatures and 
densities determined in Garnett \etal (1997b\markcite{g6}; G97b). 
However, we measured an electron 
density of 600$\pm$200 cm$^{-3}$ from the [S II] $\lambda$6717/$\lambda$6731 
ratio in our FOS spectrum of VS 44, which we 
adopted. This results in a 9\% higher O$^+$ abundance than for $n_e$ 
= 100 cm$^{-3}$, due to the effects of collisional de-excitation on 
the [O~II] 3727 \AA\ doublet. 

Recent spectroscopy of the M101 regions is not available, so we relied 
on previously published results. We found four papers (Rayo, Peimbert, 
\& Torres-Peimbert 1982\markcite{r1}; Skillman 1985\markcite{s4}; 
McCall, Rybski, \& Shields 
1985\markcite{m5}; and Torres-Peimbert, Peimbert, \& Fierro 
1989\markcite{tpf89}) in which 
[O~III] $\lambda$4363 has been measured for our targets, with 
uncertainties quoted. A five-level atom program (De Robertis, Dufour, 
\& Hunt 1987\markcite{d1}) was then used to derive $T_e$ from the 
[O~III] line measurements in these references, and the results 
averaged with weights determined from the errors. We did detect
$\lambda$4363 in our FOS spectrum of NGC 5471; the resulting $T_e$ 
derived from this spectrum was 13300$\pm$1200 K, consistent with the 
published values. For the M101 regions we have no direct measurements 
of [O~II] 7320-30 \AA, so we derive T[O~II] from our T[O~III] 
values following the formulation in Garnett (1992)\markcite{g1}; 
the resulting temperatures 
are listed also in Table 3. We derived electron densities 
from [S~II] line ratios in the four references cited above; all 
were consistent with the low-density limit, and for the abundance 
analysis we adopted $n_e$ = 100 cm$^{-3}$.

\subsection{Ionic Abundances}

We follow the prescriptions described in G97a,b to derive ionic 
abundances from the FOS spectra, using the physical conditions 
listed in Table 3. As in G97b, we employ a two-zone model for the 
temperature structure of each region to account for differences in 
cooling between the high- and low-ionization zones. Thus, C$^{+2}$, 
O$^{+2}$, Si$^{+2}$, and Ne$^{+2}$ are characterized by the [O~III] 
electron temperature, while C$^+$, N$^+$, O$^+$, and S$^+$ are 
characterized by T[O~II]. 

Given the physical conditions in Table 3, we computed level 
populations and line emissivities $\epsilon$($\lambda$) using 
the five-level atom calculation. The ionic ratio then follows from  

\begin{equation}
{{X^{+i}}\over {H^+}} = 
{\epsilon(H\beta)\over \epsilon(\lambda_i)} {I(\lambda_i)\over I(H\beta)} .
\end{equation}

The ionic abundances derived thus are listed in Table 4; two sets of 
abundances are shown, corresponding to the two cases of interstellar 
reddening we consider.

\subsection{Ionization Corrections and Final Element Abundances}

Since we observe both C~II] and C~III] as well as [O~II] and [O~III], 
we expect that any contributions to the total C and O abundances from 
unseen ionization states to be very small. Nevertheless, the most 
metal-poor, high ionization \hii\ regions in our sample, such as 
NGC 5471 with X(O$^{+2}$) $\equiv$ O$^{+2}$/O = 0.76, might have a 
non-negligible contribution from C$^{+3}$. Therefore, 
we have examined models for the photoionization of carbon 
in \hii\ regions to estimate the size of such a contribution. These 
models are as described in G95 and G97a. In brief, the models covered 
the ranges 0.1-1.0 solar O/H, 35,000 K $\le$ $T_{eff}$ $\le$ 50,000 
K in stellar temperature, and $-$4 $\le$ log $U$ $\le$ $-$2 in 
ionization parameter.

The outcome of the modeling is displayed in Figure 1, which shows 
how the summed ionization fractions of C$^+$ and C$^{+2}$ varies 
with oxygen ionization fraction X(O$^+$). Figure 1 shows that the 
contribution of C$^{+3}$ to the total carbon abundance is $<$ 10\% 
for a wide range of nebular ionization, and is non-negligible only 
for stellar temperatures 
$>$ 45,000 K. Only for X(O$^+$) $<$ 0.2 can the correction 
for C$^{+3}$ exceed 0.1 dex. NGC 5471, with X(O$^+$) = 0.24, 
has a predicted correction of at most about 10\%; for the other 
regions, the correction to the carbon abundance is of order 
5\% or less. We therefore conclude that the uncertainty in 
the ionization correction for C$^{+3}$ is small compared to the 
other sources of error for our \hii\ region sample. 

For the \hii\ region NGC 2403-VS 38 we measured only an upper limit to 
the C~III] line flux, and consequently only an upper limit for the 
C$^{+2}$ abundance can be derived. However, we can use the derived 
C$^+$/O$^+$ ratio plus the derived X(O$^+$) to estimate the C$^{+2}$ 
fraction. Figure 2 shows the variation of X(C$^+$)/X(O$^+$) vs X(O$^+$) 
from the ionization models; note that X(C$^+$)/X(O$^+$) has a small 
range of values at fixed X(O$^+$). VS 38 has X(O$^+$) = 0.54; for this 
value X(C$^+$)/X(O$^+$) is 0.8$\pm$0.1. Thus, we expect X(C$^+$) to be 
0.43$\pm$0.05. We therefore predict C$^{+2}$/H$^+$ to be 10$\times$10$^{-5}$ 
for an $R_V$ = 3.1 reddening curve, or 7$\times$10$^{-5}$ for $R_V$ = 5 
reddening. Both values are consistent with the upper limits derived 
directly from the C~III] line flux limits, and we adopt the observed
C$^+$/O$^+$ $\times$ ICF for VS 38. Likewise, for NGC 5471 we have 
only an upper limit for C~II]. For X(O$^+$) = 0.24, our models predict 
C$^+$/C$^{+2}$ $\approx$ 0.27. This is consistent with our 2$\sigma$ 
upper limit of 0.25, given the uncertainty 
in the C$^{+2}$ abundance. (At $T_e$ = 13,000 K, we expect 
[O~III] $\lambda$2322/H$\beta$ $\approx$ 0.02, about 25-30\% of our 
2$\sigma$ upper limit for C~II]. For the other \hii\ regions, [O~III] 
2322 \AA\ emission is a negligible contribution to the measured C~II] 
line strengths.) Thus, we are confident that our C/O ratios in all cases 
are representative. The total derived abundances for carbon and oxygen 
are listed in Table 5. 

\subsection{Uncertainties}

The uncertainties in the final C/O abundance ratios were estimated by 
summing in quadrature the contributions from the line flux measurements 
(including uncertainty in the differential reddening), and the differential 
uncertainties in the C~II], C~III], [O~II], and [O~III] emissivities due 
to errors in T$_e$; the small uncertainty in the ionization corrections 
for carbon is neglected. 

We consider possible sources of systematic error below.

(1) {\it Errors in electron temperature}. A systematic error in T$_e$ 
can affect the derived C$^{+2}$/O$^{+2}$ and C$^+$/O$^+$ ratios, because 
of the large difference in excitation. C$^{+2}$/O$^{+2}$ from the 
$\lambda$1909/$\lambda$5007 ratio varies as $e^{4.65/t}$ (where t = 
T$_e$/10$^4$ K), while C$^+$/O$^+$ from $\lambda$2325/$\lambda$3727 
varies as $e^{2.34/t}$. Peimbert (1967)\markcite{p1} showed that 
significant fluctuations in electron temperature about 
the average nebular value can lead one to systematically 
underestimate the abundances from collisionally-excited optical/UV 
lines, because the forbidden line emission is weighted toward 
regions with higher than average $T_e$. If this is the 
case, the true C/O ratios in the \hii\ regions could be systematically 
higher than our derived values. We do not expect this to affect 
the relative C/O values much, given the small derived values for 
such fluctuations in Galactic \hii\ regions (Esteban et al. 1998),
and given that there is no evidence that the postulated temperature 
fluctuations depend on metallicity and galaxy environment.

(2) {\it Depletion onto grains}. G95 discussed depletion of C and O 
onto grains at length. Since then, new absorption-line measurements of 
C and O in the local diffuse ISM have shed additional light on the 
subject. Meyer, Jura, \& Cardelli (1998\markcite{m6}; MJC) provide 
additional O abundances from weak 
O~I absorption lines, while Sofia \etal (1997\markcite{s2}; SCGM) have 
added new data on carbon abundances from weak C~II] absorption lines. 
The sightlines in both studies were chosen to sample a wide range of 
physical conditions. In particular, they chose sightlines with very 
low fractions of molecular hydrogen; refractory elements generally show 
a trend of decreasing gas-phase abundance with increasing $f(H_2)$, 
indicating an increasing fraction of the element incorporated into dust 
grains. MJC and SCGM, in contrast, found that neither O nor C abundances 
in the local neutral gas vary with $f(H_2)$ or average gas density, but 
rather have very similar values toward all lines of sight that they 
observed. The implication from both studies is that the C and O in the 
diffuse gas reside in resilient grains with little exchange of C and O 
between gas and dust. At the same time, the dependence of element 
depletions on metallicity is completely unknown. A study of element 
depletions for the low metallicity gas in, for example, the Magellanic 
Clouds is highly desirable.

The amount of C and O in grains depends on the choice of reference 
abundances, as discussed in Mathis (1996). Various arguments (Sofia,
Cardelli, \& Savage 1994\markcite{s3}; Meyer et al. 1994\markcite{mjhc94}; 
Mathis 1996\markcite{m3}, MJC) suggest that the solar neighborhood 
B stars, with average O/H values of about 60\% of the 
solar value (e.g., Cunha \& Lambert 1994\markcite{c4}, Kilian, 
Montenbruck \& Nissen 1994\markcite{k4}), provide a more appropriate 
local abundance reference for oxygen than the solar-type stars. If 
we adopt this argument for O and C for our \hii\ region data, then 
the results of MJC and SCGM imply that our O abundances should be 
increased by approximately 0.1 dex and our C abundances by 0.2-0.3 
dex to account for atoms locked up in grains. Esteban et al. 
(1998)\markcite{e1} also argue for about 0.08 dex of oxygen in 
grains, based on the depletion of Fe, Si, and 
Mg in the Orion Nebula. On the other hand, the energetic environment 
of giant \hii\ regions may result in 
grain destruction by some unknown amount; Calzetti \etal 
(1994)\markcite{c1} noted the general absence of the 2175 \AA\ 
feature in the spectra of starburst 
galaxies, and have argued that this may be the result of grain 
destruction. On the other hand, we clearly see the 2175 \AA\ feature 
in our spectra, so grain destruction remains an open question. For 
this paper, we choose not 
to apply a correction for grains to our derived abundances, but the 
reader should keep in mind that the actual C/O ratios may be 
higher by 0.1-0.2 dex for both the spiral and the dwarf galaxy data. 

\section{Discussion}

\subsection{The Variation of C, N, and O in NGC 2403 and M101}

Figure 3 displays the derived C/O ratios for the NGC 2403 and M101 \hii\ 
regions as a function of O/H. The open symbols in this figure show C/O 
for an $R_V$ = 3.1 reddening law; filled symbols show the $R_V$ = 5 
case. Despite the uncertainty in the reddening law, the data clearly 
show that C/O increases with O/H in the two spiral galaxies. The actual 
rate of increase can not be determined precisely without better knowledge 
of the reddening corrections and depletion factors, although the observed 
trend in C/O is at least as steep as the one derived by G95. When the 
Orion Nebula and solar abundances are included, there is some 
indication that C/O levels off at higher metallicities. 

Our new results show clearly an increase in C/O with O/H as had been 
noted by G95 from observations of \hii\ regions in irregular galaxies. 
This is demonstrated graphically in Figure 4, 
where we plot the spiral galaxy data together with published data 
for irregular galaxies. The spiral galaxy points merge smoothly 
with and extend the trend over the range $-$4.0 
$<$ log O/H $<$ $-$3.3. At present it is not 
possible to tell if the trend in C/O flattens at lower O/H (in 
which case I~Zw~18 could be considered to have normal C/O), or if 
C/O continues to decline at lower O/H (in which case I~Zw~18 would 
have unusually high C/O). Additional data on C/O in metal-poor 
irregular galaxies are needed to clearly show if C/O flattens at the 
highest and/or lowest metallicities.

G95 also noted an apparent trend of increasing C/N with O/H in the 
irregulars, with an abrupt decline in C/N for the Orion nebula and 
solar neighborhood stars, and suggested that the trend could reflect 
the differences in star formation histories between spirals and 
irregulars. Our new spiral galaxy data suggest a considerably 
different picture, as shown in Figure 5. With our new data added, 
no significant trend is seen in C/N vs. 
O/H, either in the combined data, or in the spiral galaxy data 
considered separately. Even if we exclude the low points for 
NGC 5253, which may be contaminated by N-enriched Wolf-Rayet star 
ejecta (Kobulnicky et al. 1997\markcite{k7}), no correlation is 
evident. It may be, therefore, that the trend in C/N noted by G95 
was the result of a limited number of data points. The absence
of a trend in C/N indicates that the two elements are injected
into the ISM on similar timescales.

\subsection{Evolution of C/O vs. O/H: Solar Neighborhood Models}

The interpretation of the trend of carbon abundances involves an 
extra level of complexity, because carbon can be synthesized in 
long-lived intermediate mass stars as well as massive stars. Tinsley 
(1979)\markcite{t3} showed that a primary element, such as C, that 
is produced in long-lived stars can mimic the behavior of a secondary 
element, because of the delay in ejection of C into the ISM. As a 
result, the instantaneous recycling approximation can not be applied 
for carbon. Thus, one must either model the evolution numerically, 
or model the delayed ejection with an analytic approximation as in 
Pagel (1989)\markcite{bejp89}.

In G95, we compared the observed trend for C/O with O/H in irregular
galaxies with the results of chemical evolution computations for the
evolution of C/O in the solar neighborhood. On the basis of those 
models, we concluded that the observed variation in C/O was best 
explained with the metallicity-dependent yields for carbon and oxygen 
derived for massive stars with radiatively-driven mass loss (Maeder 
1992\markcite{m1}), plus the intermediate mass star contribution. 
Models which used hydrostatic massive star model yields failed to 
predict a steep enough increase in C/O at high O/H. Here we re-examine 
the comparison of models with our new results.

Figure 6 shows C/O vs. O/H for the combined spiral and irregular galaxy
sample, along with a few representative models for the evolution of C/O 
in the solar neighborhood. Figure 6(a) shows models by Carigi 
(1994\markcite{c2}, 1996\markcite{c3}), using massive star yields from 
Maeder (1992)\markcite{m1}, which include the effects of mass loss on 
C and O yields; Figure 6(b) shows the predictions of models by Timmes, 
Woosley, \& Weaver (1995)\markcite{t2} and Chiappini, Matteucci, \& 
Gratton (1997)\markcite{cmg97}; both use the massive star yields of 
Woosley \& Weaver (1995)\markcite{w2}, which do not include the 
effects of stellar mass loss. All of the models shown use Renzini \& 
Voli (1981)\markcite{r2} as the source for intermediate mass star 
yields for carbon and other elements. 

The effect of the choice of massive star yields is clear comparing Fig. 
6(a) with Fig. 6(b): the models with Maeder yields show steeper increases 
in C/O at high metallicities. (The models of Prantzos, Vangioni-Flam, \& 
Chauveau 1994\markcite{p2} show similar behavior.) Do the data favor one 
family of models? Unfortunately, 
the answer depends on the assumed UV reddening function. When 
the flatter UV reddening is used, the data are consistent with the models 
using WW95 yields. On the other hand, with the steeper $R_V$ = 3 reddening 
function the data are more consistent with the models using Maeder's 
massive star yields. Improved UV spectra to determine the proper reddening 
correction for metal-rich \hii\ regions are needed to distinguish clearly 
between the two families of chemical evolution models.

At the same time, solar neighborhood chemical evolution models may not
provide a unique representation of the evolution of C/O vs. O/H. For 
example, compare the Carigi (1996)\markcite{c3} model in Fig. 6(a) 
with the Carigi (1994)\markcite{c2} model. Both use the same yields, 
yet the abundance evolution is noticeably different. 
One explanation for the differences is that Carigi (1996)\markcite{c3} 
uses a star formation law with a steeper dependence on 
gas surface density than Carigi (1994)\markcite{c2}. As a result, in the 
Carigi (1996)\markcite{c3} model star formation and enrichment progress 
more slowly; because of the slower evolution, carbon production from 
lower mass stars has a chance to ``catch up'' more quickly with the 
oxygen production, and so C/O is higher in the 1996 models at low O/H. 
This may also account for the difference between the Timmes et al. and 
Chiappini et al. models, since Timmes et al. use a steeper star formation 
law than Chiappini et al. One can infer from this exercise that the 
evolution of C/O with O/H is sensitive to the star formation/enrichment 
timescale. Therefore, solar neighborhood models are not likely to predict 
correctly the evolution of abundance ratios in regions with different gas 
consumption timescales.

What about the abundances as a function of the gas mass fraction $\mu_g$? 
Figure 7 shows the abundance ratios O/H and C/O as a function of ln $\mu_g$. 
Gas fractions for NGC 2403 were taken from G97b. For M101, we used H~I 
measurements from Kamphuis (1993)\markcite{k1}, CO measurements from 
Kenney, Scoville, \& Wilson (1991)\markcite{k2}, and surface photometry 
from Okamura, Kanazawa, \& Kodaira (1976)\markcite{o1} to construct the 
gas fraction profile, assuming that NGC 2403 and M101 have the same 
stellar mass/light ratio; we also assumed for convenience a symmetric 
gas profile for 
M101, although the observations of Kamphuis show that the H~I surface 
density is not symmetric about the galaxy's minor axis. We do not use 
dynamical models to determine masses because of the uncertain effect of 
dark matter on the comparison between models and observations. Although 
it is difficult to estimate uncertainties precisely for H$_2$ column 
densities derived from CO measurements and for disk M/L ratios, we assign 
a provisional uncertainty of $\pm$20\% for the derived gas fractions. 

Figure 7 shows predictions of abundances vs. gas fraction for the models 
of Carigi (1996)\markcite{c3} and Chiappini et al. (1997)\markcite{cmg97}. 
The Carigi model ({\it solid curve}) again shows a steep increase in C/O
due to the effects of stellar mass loss on the massive star yields. This
model provides a good qualitative match to the trend of the M101 data 
points, although it systematically overpredicts both C/O and O/H. The 
discrepancy is somewhat reduced if a modest amount (0.1-0.2 dex) of C and 
O in the ionized gas is in dust grains. On the other hand, the Chiappini 
et al. model ({\it dashed curve}) shows very flat behavior for C/O once 
the gas fraction falls below 50\% (ln $\mu$ $\approx$ $-$0.6). This model 
would be qualitatively consistent with the observations for the case of 
the $R_V$ = 5 reddening function, it would fail to explain the steeply 
rising C/O if the $R_V$ = 3 reddening function applies. The comparison
here highlights again the need for additional, improved measurements of 
UV reddening and emission line strengths.

We see marginal evidence for systematically higher C/O at a given $\mu$ 
in M101 than in NGC 2403. This may be related to the observation that
interstellar abundances are higher overall in more massive spirals (G97b).
However, with the limited amount of available data nothing definite can 
be concluded at this time about differences in C/O ratios between spirals.

\subsection{Radial Gradients in C/O and Chemical Evolution Models}

Another key test of the chemical evolution models is whether they 
reproduce the spatial distribution of heavy elements across the galaxies. 
Radial gradients in the oxygen abundance are commonly observed in spirals, 
and the various published chemical evolution models generally reproduce 
the O/H gradients. Now we have information on the spatial distribution 
of carbon in two galaxies, which potentially provides additional 
constraints on the models. The questions are whether we see gradients 
in C/O in these galaxies, and whether existing chemical evolution 
models reproduce the observed trends.

Before we can discuss these questions, it is necessary to determine how 
best to compare the models with the data for NGC 2403 and M101. Few 
theoretical studies have discussed the spatial distribution of carbon 
in spirals, presumably because of a lack of measurements to compare with 
the models. The three studies which do provide spatial distributions for 
carbon (Carigi 1996\markcite{c3}, G\" otz \& K\" oppen 1992\markcite{g7}, 
and Moll\' a et al. 1997\markcite{m7}) present results based on models 
for the Galaxy. Moll\' a et al. presented models for a few other spirals 
(not for NGC 2403 or M101, however), but the variation in their predicted 
C/O gradients is small. Direct comparison of model abundance gradients, 
in dex/kpc, for the Milky Way with observations of other spirals is not 
informative, because spirals come in a variety of sizes. 
However, G97b showed that high surface brightness spiral galaxies 
have similar oxygen abundance gradients when plotted per 
unit scale length, with no correlation with galaxy luminosity. This 
offers a potential way to scale the Milky Way models for 
comparison with other galaxies, and we do so here. For the purposes of 
our discussion, we adopt the following disk scale lengths: R(scale) = 
2.1 kpc for NGC 2403 (based on the photometry of Okamura, Takase, \& 
Kodaira 1977\markcite{o2}, 
plus the Cepheid distance of Tammann \& Sandage 1968\markcite{st68}); 
R(scale) = 3.5 kpc for the Galaxy (de Vaucouleurs \& Pence 
1978\markcite{dp78}); and 5.4 kpc for M101 (Okamura et 
al. 1976\markcite{o1}, plus the Cepheid distance of Kelson et al. 
1996\markcite{km101}). We then normalize galactocentric distances in 
the observational sample and in the 
chemical evolution models by R(scale).

We plot the abundance data for NGC 2403 and M101 as a function of the 
normalized radius R/R(scale) in Figures 8 and 9, with the Milky Way 
chemical evolution models overplotted. Note first of all that the 
observations imply that both NGC 2403 and M101 show significant radial 
gradients in C/O, of order $-$0.1 to $-$0.2 dex/scalelength.  

The upper panels of Figs. 8 and 9 show that all of the chemical evolution 
models considered here do an adequate job of reproducing the slope of the 
abundance gradients, when normalized to the disk scalelength. The models 
tend to predict significantly higher O abundances at a given R/R(scale) 
than observed in NGC 2403. This may reflect a real difference in the 
global enrichment. G97b noted that the most luminous spirals have higher
abundances at a fixed value of disk surface brightness than their less 
luminous counterparts. This reflects the relationship between galaxy 
mass and metallicity observed in spiral galaxies (Garnett \& Shields 
1987\markcite{g2}, Zaritsky, Kennicutt, \& Huchra 1994\markcite{z1}). 

The lower panels in Figs. 8-9 compare the model calculations for the 
radial variation of C/O with our data. In contrast to the case of O/H, 
the models vary widely in their predictions for the variation of C/O. 
The model of Carigi (1996)\markcite{c3} in Figure 8 appear to be an 
adequate match for the M101 data, although the model somewhat overpredicts 
both C/O and O/H. This discrepancy is reduced if a modest amount of C 
and O are in dust grains. Alternatively, the derived O/H could be too
low because of temperature fluctuations (Peimbert 1995\markcite{p95}), 
or perhaps the formation of black holes by massive stars above a certain 
mass reduces the stellar yields (Maeder 1992\markcite{m1}).
The systematically lower abundances in NGC 2403 await an explanation; 
this may be related to the question of what mechanism is responsible 
for the mass-metallicity relation for galaxies. On the other 
hand, the models of G\" otz \& K\" oppen (1992)\markcite{g7} and 
Moll\' a et al. (1997)\markcite{m7}, shown in Fig. 9, produce C/O 
gradients which are too shallow compared to the data. G\" otz \& 
K\" oppen, in fact, predict a slight {\sl increase} in C/O progressing 
outward. Moll\' a et al. actually predict a range in C/O gradients for 
the galaxies they modeled, 0.0 to $-$0.01 dex/kpc; here we used their 
steepest predicted gradient, computed for the Milky Way.

Curiously, the models in Figure 9 do not fail by producing too little 
C in the inner disks, as one might have expected since both models 
use Santa Cruz massive star yields computed with without stellar mass 
loss. Instead, both models predict too much carbon in the outer disks. 
Note, however, that G\" otz \& K\" oppen artificially increased the 
yield of carbon in their model by a factor of three, motivated by 
the failure of their model to reproduce the solar carbon abundance. 
A uniform reduction in their C abundances by a factor three would 
bring their outer disk abundances in line with our observed C/O in 
NGC 5471, and then they would underproduce C in the inner disks, as 
expected. On the other hand, the implied high C/O in the outer disks 
of the Moll\' a et al. models is more difficult to understand. It 
seems clear that explaining the heavy element abundance pattern in 
the outer disks of spiral galaxies remains a challenge to theoretical 
models.

\section{Conclusions}

Because of the unique ultraviolet spectroscopic capability of the {\it 
Hubble Space Telescope}, we are able to present the first significant
sampling of the spatial distribution of carbon abundances in 
spiral galaxies. We have shown that there are radial gradients in C/O 
across the disks of both NGC 2403 and M101, in the sense that the C/H 
gradients are steeper than the gradients in O/H. On the other hand, 
our ability to determine the actual magnitude of the C/O gradients is 
limited because of the uncertainty in the choice of UV reddening function. 
C/N appears to show no correlation with O/H, as might be expected if
both are produced in similar stellar mass ranges. Again, however, 
more and improved data are needed to reduce some of the scatter seen
in Figure 5, to more precisely constrain systematic trends in C/N.

The next observational step will be to obtain improved UV spectroscopy 
with sufficient signal/noise in the UV He~I lines to constrain the 
reddening, as well as improved signal/noise in the C~III] and C~II] 
lines themselves, and to possibly detect N~III] 1750 \AA. (The best 
approach would be to observe O~III] 1661-6 \AA\ together with C~III], 
which would minimize the uncertainties due to errors in reddening 
and temperature; however, such observations may not be possible with 
existing HST spectrographs.) In addition, 
it is well established that abundance gradient 
determinations based on only a few data points can be very 
inaccurate (Zaritsky, Kennicutt, \& Huchra 1994\markcite{z1}), and 
spiral galaxies may show asymmetries in the abundance distribution 
(e.g., M101: Kennicutt \& Garnett 1996\markcite{k3}). It will be 
necessary, therefore, to obtain observations of additional regions to 
accurately define the radial distribution of carbon in these galaxies.

At the same time, it is clear that additional theoretical work is needed 
to understand the evolution of carbon in galaxies. Few theoretical studies 
so far have addressed the evolution of carbon, yet carbon is an important 
diagnostic of both stellar yields and the enrichment timescales. One 
important aspect to understand is the problem of the abundance pattern 
in outer spiral disks. The CNO abundance ratios in outer disks are very 
similar to those observed in dwarf irregular galaxies (see also Ferguson 
et al. 1998\markcite{f1}, van Zee et al. 1998\markcite{v1}). This might
be difficult to understand in the context of a disk which is more or 
less uniformly old, but has had the outer disk abundances modified by 
slow accretion of metal-poor gas from the outer galaxy. The implication
is that outer spiral disks have experienced slow star formation much 
like the dwarf irregular galaxies, and that the average age of stars
in outer spiral disks is much smaller than in the inner disk (cf. Allen,
Carigi, \& Peimbert 1998\markcite{acp98}). This is consistent with the 
concept of ``inside-out'' build-up of the disk, in which the timescale 
for accretion of gas onto the forming disk increases radially outward
(e.g., Chiappini et al. 1997\markcite{cmg97}). 

One important question is whether the similarity of outer spiral disk 
abundances to the irregular galaxies is a general phenomenon, or whether 
there are significant differences between late-type spirals and early-type 
spirals. We have obtained only a small number of data points in two
Sc-type spirals, barely the tip of an iceberg. Additional, improved 
measurements of abundances of carbon and nitrogen will provide new 
and interesting information on similarities and differences in the
evolution of spiral disks.

\acknowledgments
We thank Denise Taylor and Tony Keyes at STScI for their very helpful 
assistance in implementing this program; Tony also kindly provided the 
updated FOS flat fields upon request. We also thank Rene Walterbos for 
providing ground-based images for some of our targets, and Michael Rosa 
for supplying coordinates for his earlier FOS observations of the M101 
\hii\ regions. DRG and MP thank Leticia Carigi and Cristina Chiappini 
for extensive and informative discussions of the chemical evolution 
models and for providing tables of numerical results. Support for this 
program was provided by NASA and STScI through grant GO-6044-94A. DRG 
also acknowledges support from NASA-LTSARP grant NAG5-6416, while EDS 
acknowledges support from NASA-LTSARP grant NAGW-3189.

\clearpage

\begin{figure}
\plotfiddle{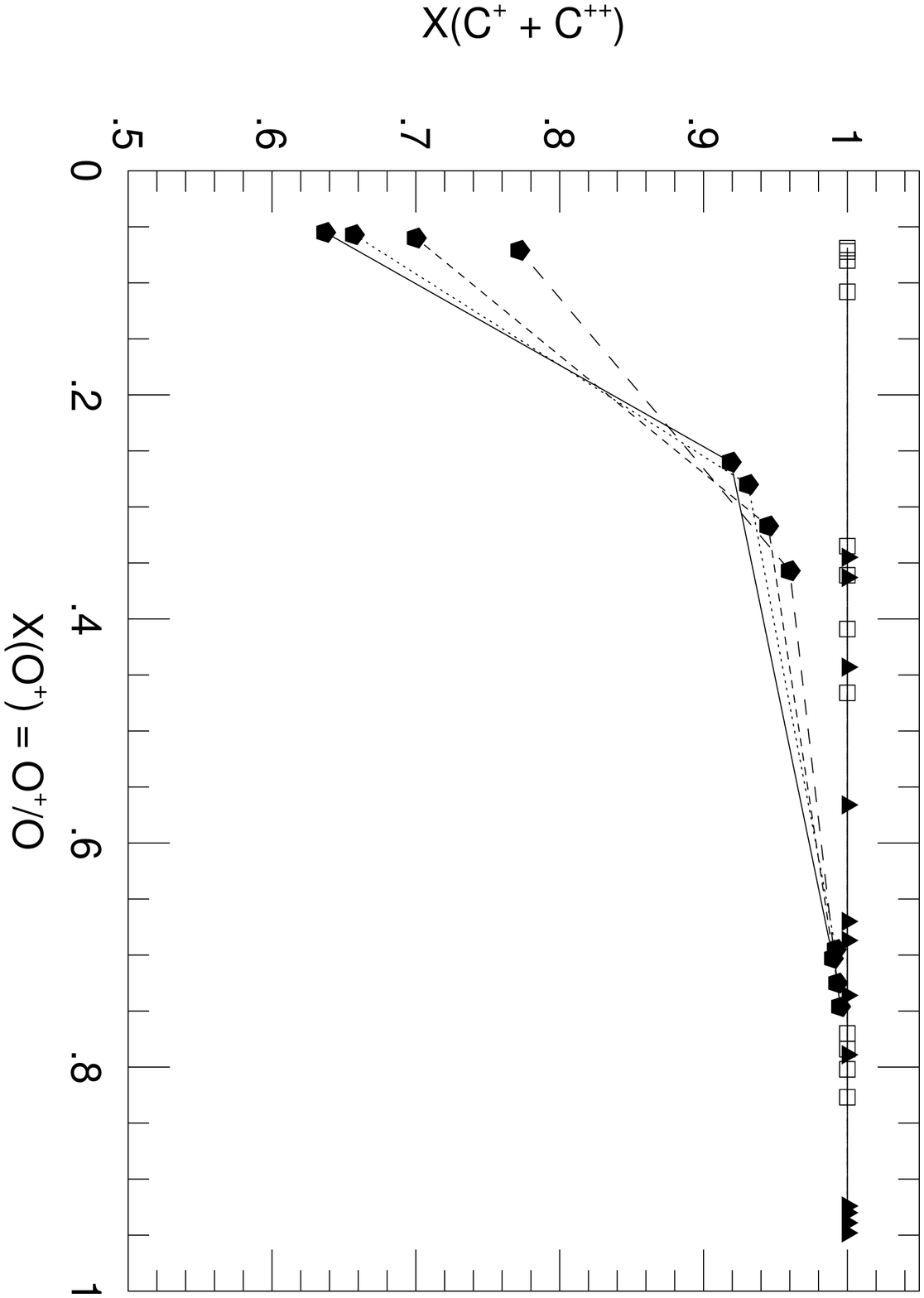}{7in}{90}{70}{70}{290}{100}
\caption{Photoionization model calculations for the ionization of carbon 
and oxygen in H II regions. The plot shows the volume fraction of C in 
C$^+$ and C$^{+2}$ versus the volume fraction of O in O$^+$. Filled 
triangles represent models with T$_{eff}$ = 35,000 K, open squares 
T$_{eff}$ = 40,000 K, and filled pentagons T$_{eff}$ = 50,000 K. Solid 
line: represents models with 0.2 times solar heavy element abundances; 
dotted line = 0.5 times solar abundances; short-dashed line = solar 
abundances; long-dashed line = 2 times solar abundances. The ionization 
parameter decreases from right to left in the plot, from log U = $-$2 
to log U = $-$4. All of the objects in our study have X(O$^+$) $>$ 0.2, 
implying negligible corrections for unobserved C$^{+3}$. }
\end{figure}

\clearpage

\begin{figure}
\plotfiddle{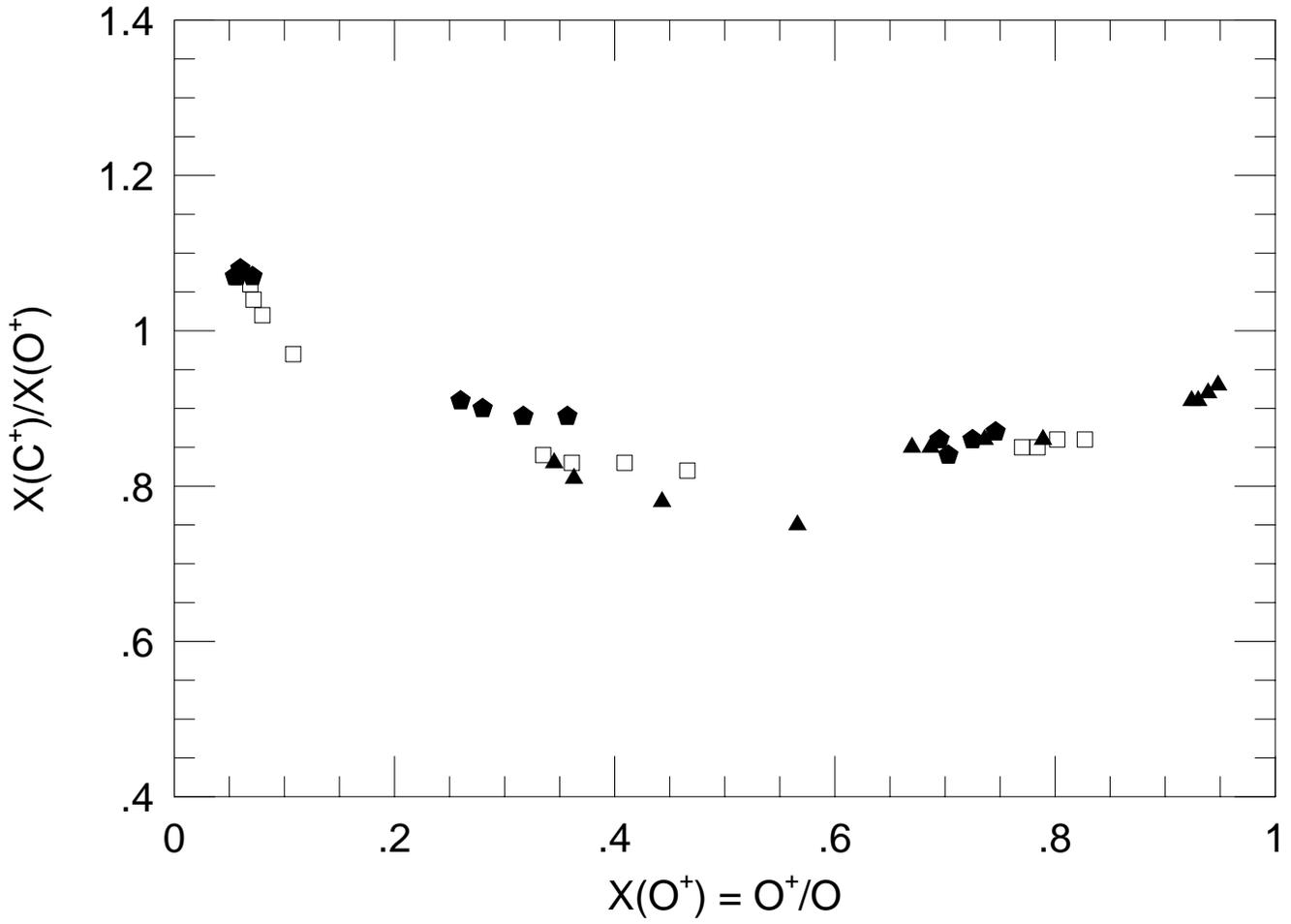}{7in}{90}{70}{70}{290}{100}
\caption{Similar to Figure 1, showing X(C$^+$)/X(O$^+$) vs. X(O$^+$), 
based on the photoionization models. }
\end{figure}

\clearpage

\begin{figure}
\plotfiddle{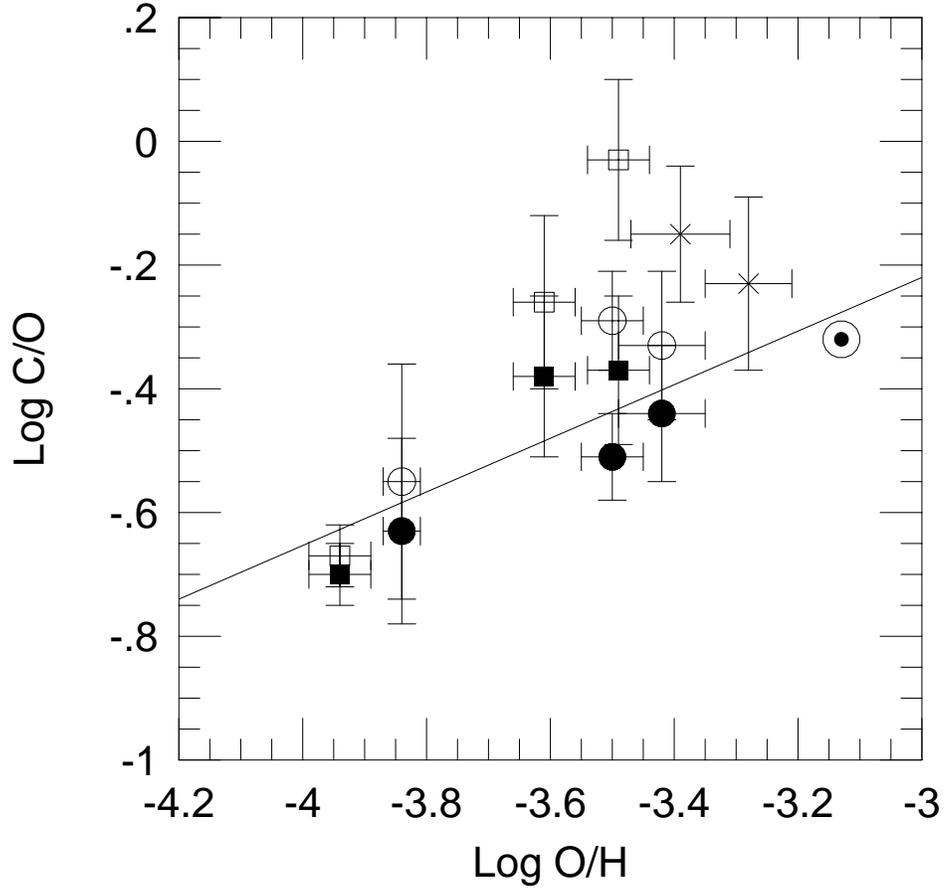}{7in}{90}{70}{70}{290}{100}
\caption{ Log C/O vs. log O/H for the NGC 2403 (circles) and M101 (squares) 
H II regions. The open symbols represent the abundances derived using a 
reddening function with $R_V$ = 3.1 to correct the spectra; filled symbols 
show the values based on an $R_V$ = 5 reddening function. Crosses are the 
values for the Orion Nebula from Walter et al. (1992) and Esteban et al. 
(1998). The line represents the relation derived by G95 for irregular 
galaxies. } 
\end{figure}

\clearpage

\begin{figure}
\plotfiddle{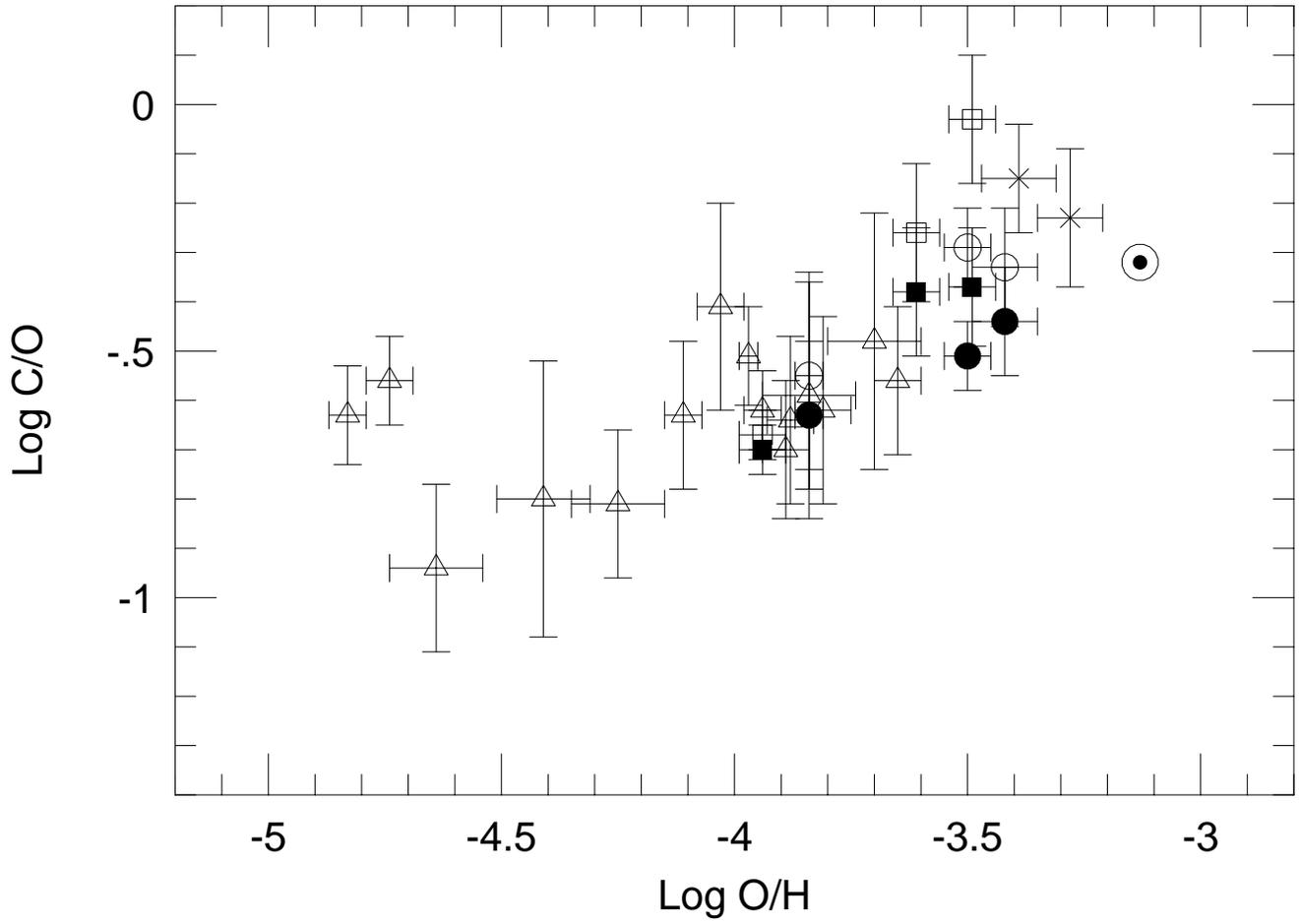}{7in}{90}{70}{70}{290}{100}
\caption{ Same as in Figure 3, but with abundance data for dwarf irregular 
galaxies and the Magellanic Clouds plotted as triangles (Garnett et al. 1995, 
1997a, Kobulnicky \& Skillman 1998). }
\end{figure}

\clearpage

\begin{figure}
\plotfiddle{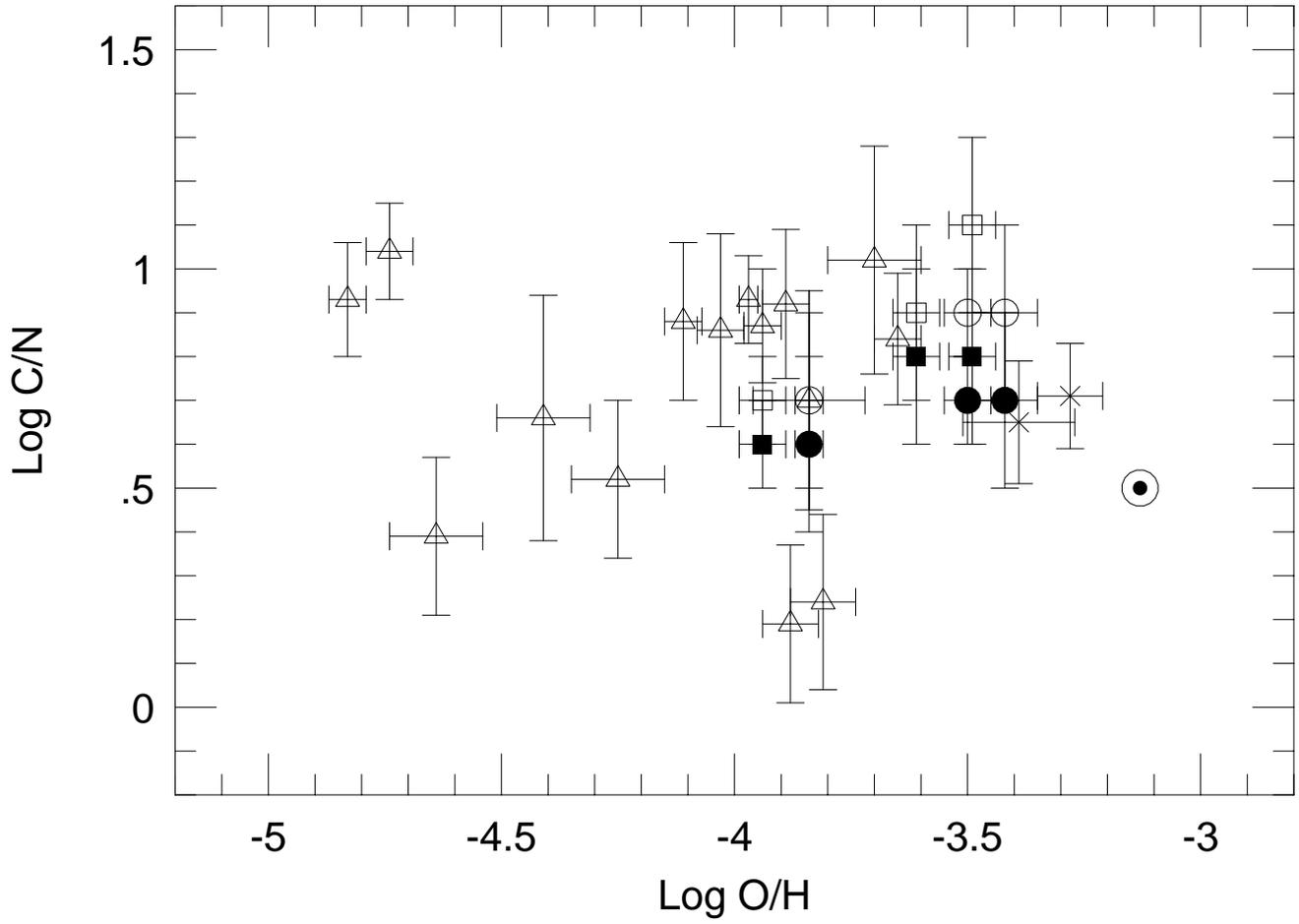}{7in}{90}{70}{70}{290}{100}
\caption{ C/N vs. O/H for the H II regions shown in Figure 4. The two
lowest points near log O/H = $-$3.85 are the N-rich regions of NGC 5253
(Kobulnicky et al. 1997).}
\end{figure}

\clearpage

\begin{figure}
\plotfiddle{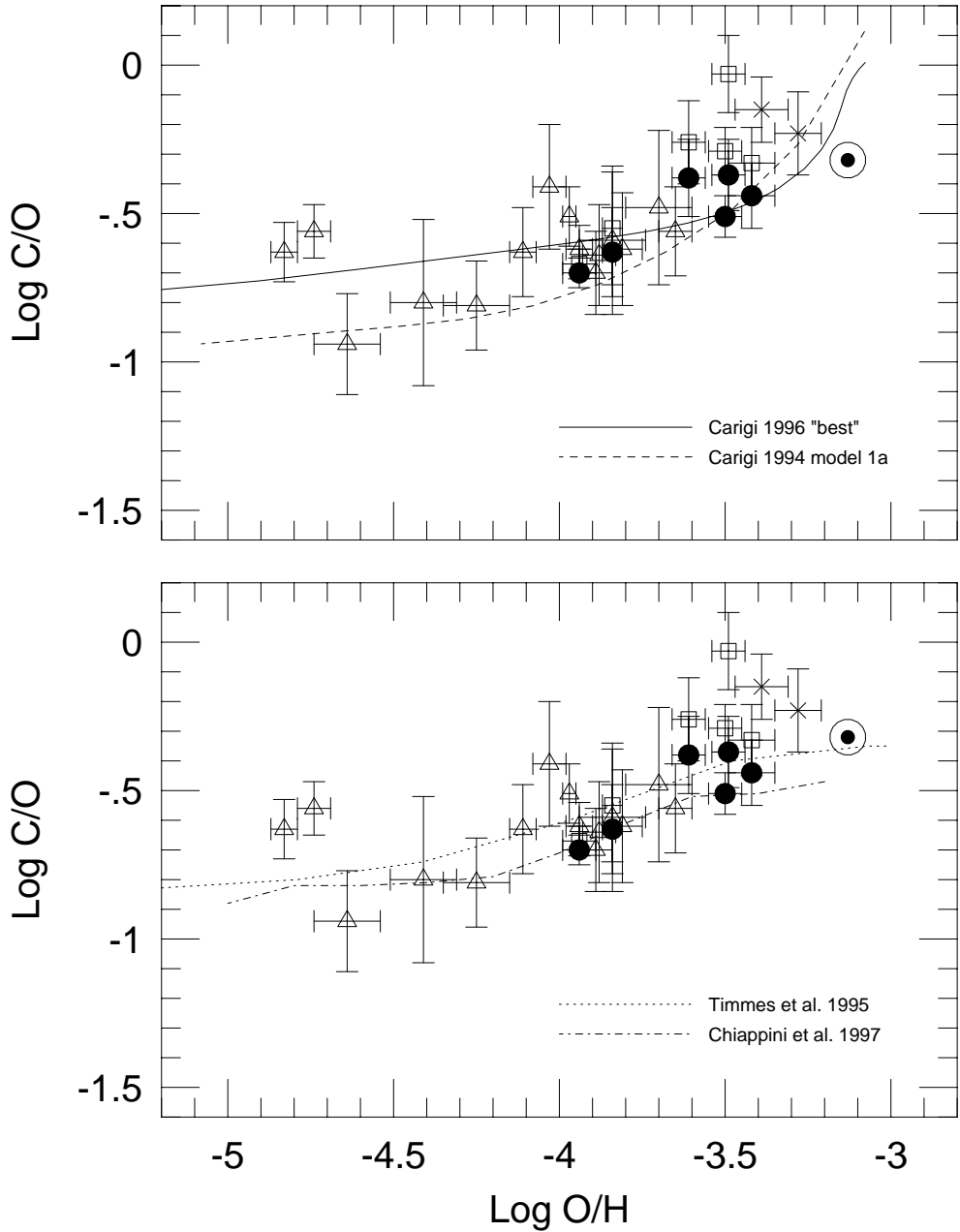}{7in}{00}{70}{70}{-220}{000}
\caption{ Carbon abundance data for spiral and irregular galaxies compared 
with models for the evolution of C/O in the solar neigbhborhood. (a) Solid 
line: the ``best model'' of Carigi (1996); dashed line: model 1a from Carigi 
(1994). These models employ the metallicity-dependent yields from Maeder 
(1992) for massive stars. (b) Chemical evolution models using massive star 
yields from Woosley \& Weaver (1995). Dotted line: Timmes et al. (1995); 
dot-dash line: Chiappini et al. (1997).}
\end{figure}

\clearpage

\begin{figure}
\plotfiddle{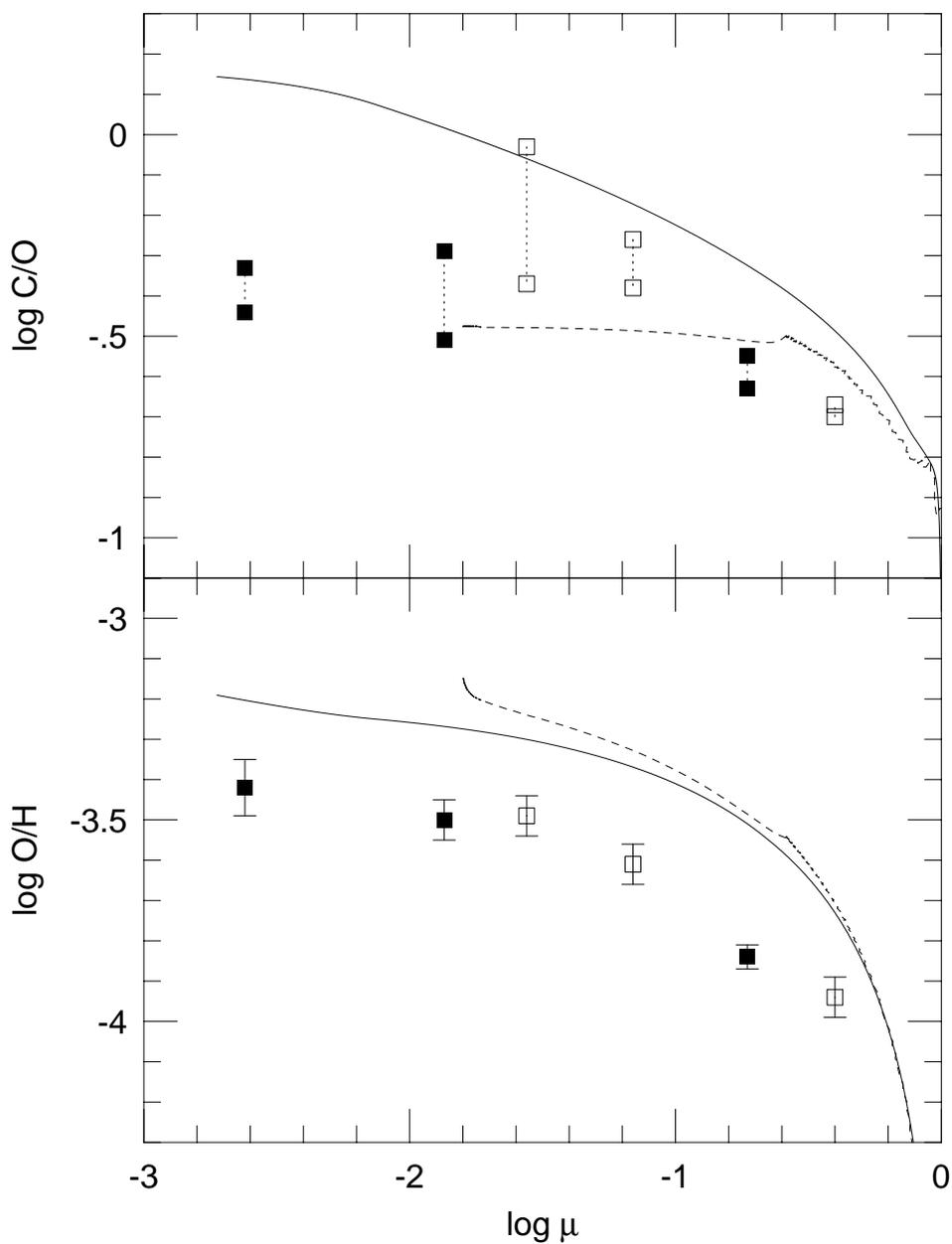}{7in}{00}{70}{70}{-220}{000}
\caption{ Abundance ratios for the NGC 2403 and M101 regions plotted
against the natural logarithm of the gas fraction. Dashed lines connect
values for the same regions, but computed with different reddening laws.
Open symbols are the M101 regions, filled symbols the NGC 2403 regions.
The solid lines show the predictions for the ``best'' model from Carigi 
(1996); the dashed line is the model of Chiappini et al. (1997). }
\end{figure}

\clearpage

\begin{figure}
\plotfiddle{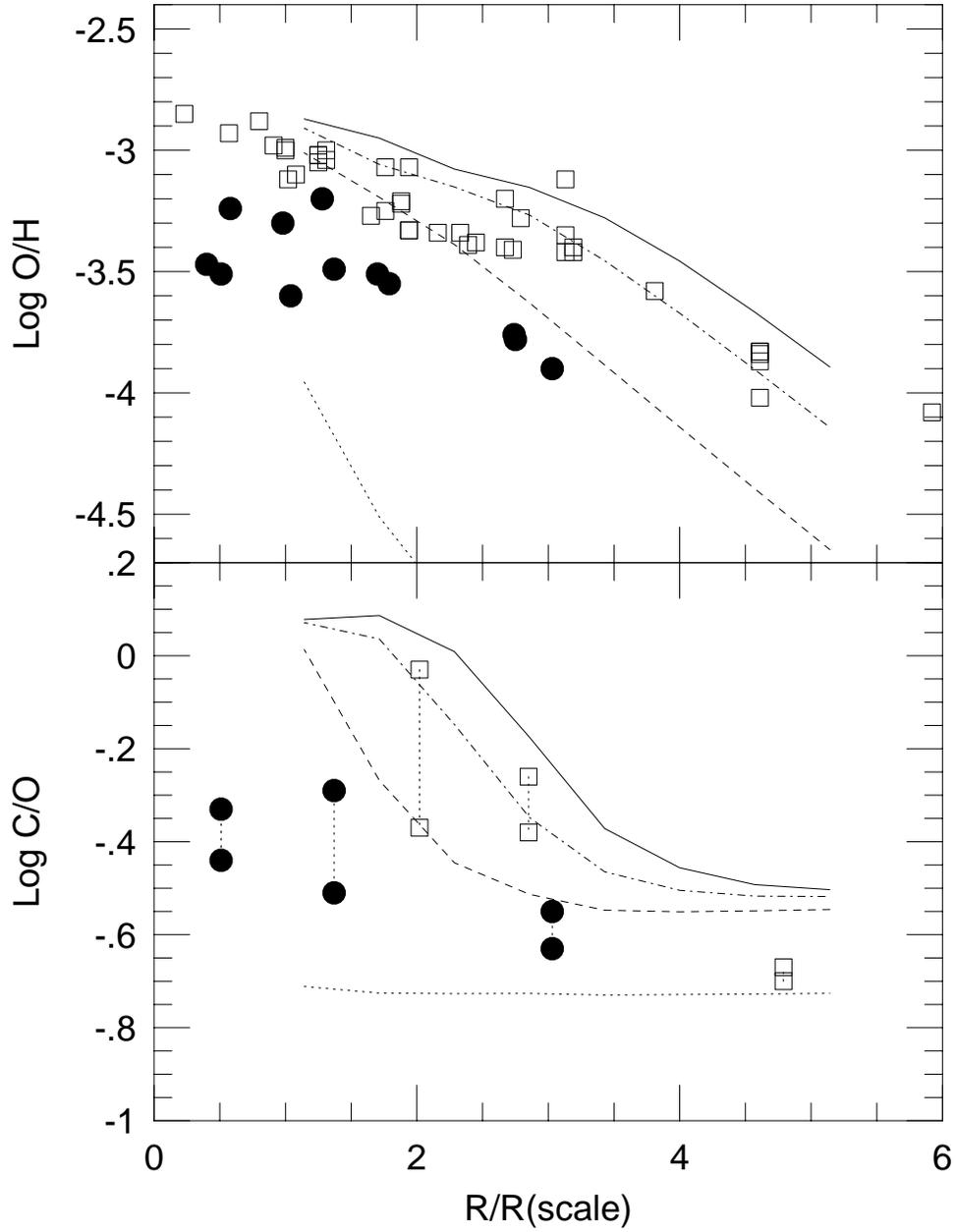}{7in}{000}{70}{70}{-220}{000}
\caption{ C and O abundances in NGC 2403 and M101 plotted as a function
of galactocentric radius normalized by the disk scale length. Open symbols:
M101; filled symbols: NGC 2403. The top panel includes all the O/H measurements
from Kennicutt \& Garnett (1996) and Garnett et al. (1997b). The lines show
model results for the Milky Way from Carigi (1996) for four different ages:
dotted line = 0.5 Gyr; dashed = 4 Gyr; dot-dash = 9 Gyr; solid = 13 Gyr. }
\end{figure}

\clearpage

\begin{figure}
\plotfiddle{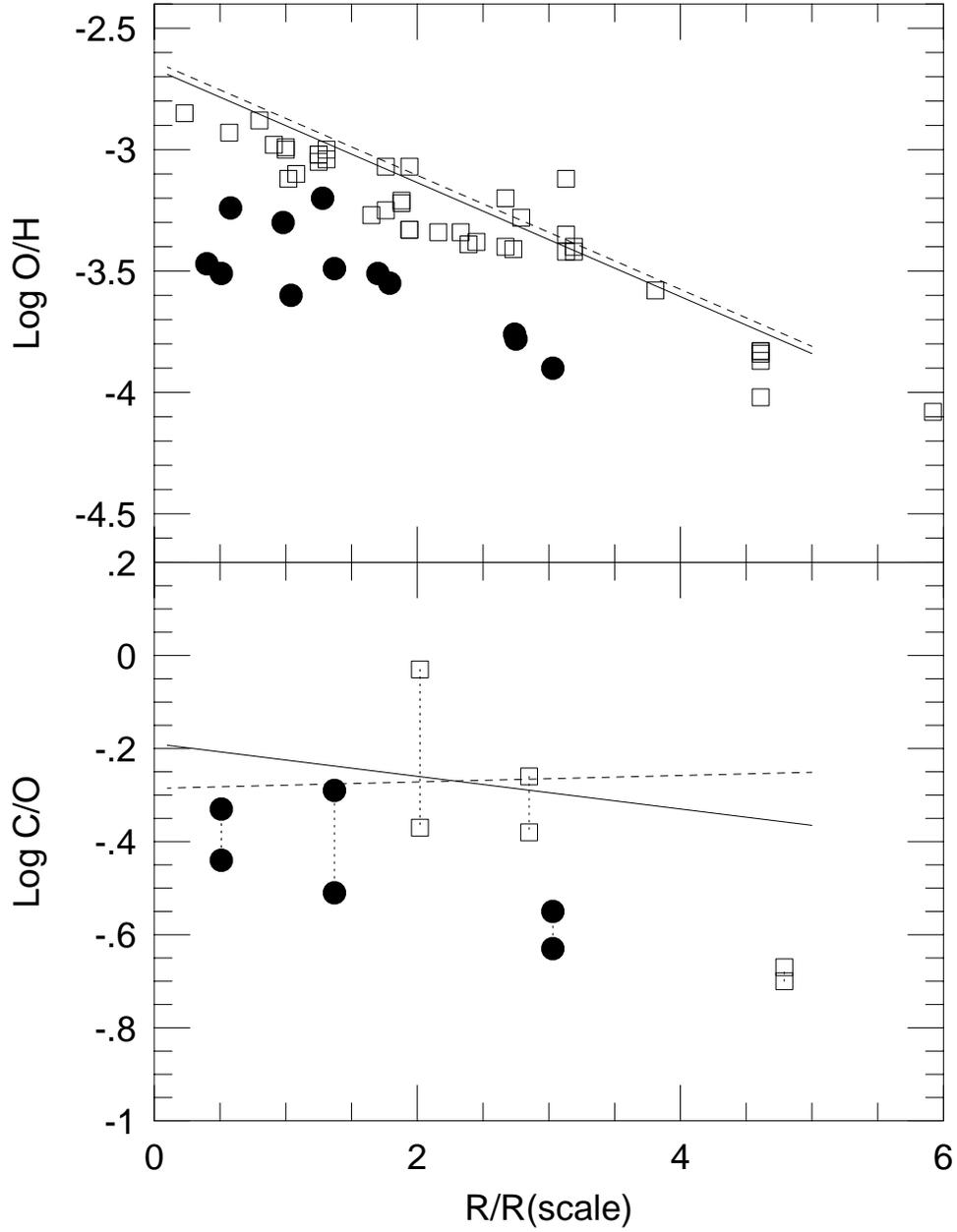}{7in}{000}{70}{70}{-220}{000}
\caption{ Same as Figure 8, but showing Milky Way model calculations
from G\" otz \& K\" oppen (1992) and Moll\' a et al. (1997). Solid line:
Moll\' a et al.; dashed: G\" otz \& K\" oppen. }
\end{figure}

\clearpage

\begin{deluxetable}{lcccccr}
\tablewidth{42pc}
\tablecaption{Journal of HST Observations }

\tablehead{
\colhead{Observation } & \colhead{Date} & \colhead{Target} & \colhead{RA} & \colhead{DEC} & \colhead{Disperser} & \colhead{Exp.} \\
\colhead{ID} & \colhead{} & \colhead{} & \colhead{(J2000)} & \colhead{(J2000)} & \colhead{} & \colhead{Time} \nl
}
\startdata
Y30A0402T & 11 Apr 1996 & M101-NGC5471 & 14:04:29.02 & +54:23:48.8 & G190H  &   1500s  \nl
Y30A0404T & 11 Apr 1996 & M101-NGC5471 & & & G190H  &   1260s  \nl
Y30A0406T & 11 Apr 1996 & M101-NGC5471 & & & G400H  &    390s  \nl
Y30A0408T & 11 Apr 1996 & M101-NGC5471 & & & G570H  &     40s  \nl
Y30A0409M & 11 Apr 1996 & M101-NGC5471 & & & G570H  &    340s  \nl
Y30A040AM & 11 Apr 1996 & M101-NGC5471 & & & G270H  &   1200s  \nl
          &             &              & & &        &          \nl
Y30A0502T & 12 Apr 1996 & M101-NGC5461 & 14:03:41.62 & +54:19:04.4 & G190H  &   1500s  \nl
Y30A0504T & 12 Apr 1996 & M101-NGC5461 & & & G190H  &   1260s  \nl
Y30A0506T & 12 Apr 1996 & M101-NGC5461 & & & G400H  &    390s  \nl
Y30A0508T & 12 Apr 1996 & M101-NGC5461 & & & G570H  &     40s  \nl
Y30A0509T & 12 Apr 1996 & M101-NGC5461 & & & G570H  &    310s  \nl
Y30A050AT & 12 Apr 1996 & M101-NGC5461 & & & G270H  &   1200s  \nl
          &             &              & & &        &          \nl
Y30A0602T & 04 Apr 1996 & M101-NGC5455 & 14:03:01.16 & +54:14:27.0 & G190H  &   1680s  \nl
Y30A0604T & 04 Apr 1996 & M101-NGC5455 & & & G190H  &   2580s  \nl
Y30A0606T & 04 Apr 1996 & M101-NGC5455 & & & G400H  &    600s  \nl
Y30A0608T & 04 Apr 1996 & M101-NGC5455 & & & G270H  &   1500s  \nl
Y30A060AT & 04 Apr 1996 & M101-NGC5455 & & & G270H  &   1080s  \nl
Y30A060CT & 04 Apr 1996 & M101-NGC5455 & & & G570H  &    300s  \nl
          &             &              & & &        &          \nl
Y30A0102T & 13 Nov 1995 & NGC2403-VS9  & 07:36:28.63 & +65:33:49.0 & G270H  &   2430s  \nl
Y30A0104T & 13 Nov 1995 & NGC2403-VS9  & & & G190H  &   3600s  \nl
Y30A0106T & 13 Nov 1995 & NGC2403-VS9  & & & G400H  &    360s  \nl
Y30A0108T & 13 Nov 1995 & NGC2403-VS9  & & & G570H  &    360s  \nl
          &             &              & & &        &          \nl
Y30A0302T & 25 Feb 1996 & NGC2403-VS38 & 07:36:52.13 & +65:36:48.5 & G190H  &   3540s  \nl
Y30A0304T & 25 Feb 1996 & NGC2403-VS38 & & & G190H  &   3540s  \nl
Y30A0306T & 25 Feb 1996 & NGC2403-VS38 & & & G570H  &    330s  \nl
Y30A0308T & 25 Feb 1996 & NGC2403-VS38 & & & G400H  &    390s  \nl
Y30A0309T & 25 Feb 1996 & NGC2403-VS38 & & & G270H  &   3840s  \nl
          &             &              & & &        &          \nl
Y30A5202T & 29 Jan 1997 & NGC2403-VS44 & 07:37:06.82 & +65:36:38.7 & G190H  &   3460s  \nl
Y30A5204T & 29 Jan 1997 & NGC2403-VS44 & & & G270H  &   2560s  \nl
Y30A5206T & 29 Jan 1997 & NGC2403-VS44 & & & G400H  &    390s  \nl
Y30A5208T & 29 Jan 1997 & NGC2403-VS44 & & & G570H  &    300s  \nl
\enddata
\end{deluxetable}

\clearpage

\begin{deluxetable}{lccc}
\tablewidth{42pc}
\tablecaption{FOS Spectra for M101 and NGC 2403 H II Regions}
\tablehead{
\colhead{ Line ID } & \colhead{ $I_{\lambda}(obs)/I(H\beta)$ } & \colhead{ $I_{\lambda}(corr)/I(H\beta)$ } & \colhead{ $I_{\lambda}(corr)/I(H\beta)$ } \\ 
\colhead{}  & \colhead{} & \colhead{$R_V$ = 3.1} & \colhead{$R_V$ = 5} \nl
}
\tablecolumns{4}
\startdata
\cutinhead{NGC 5455}                   
O III] 1666     & $<$ 0.13         & $<$ 0.25        & $<$ 0.17         \nl
N III] 1750     & $<$ 0.13         & $<$ 0.24        & $<$ 0.17         \nl
Si III] 1883    & $<$ 0.10         & $<$ 0.20        & $<$ 0.14         \nl
C III] 1909     &     0.14 (0.05)  &     0.27 (0.11) &     0.20 (0.07)  \nl
N II] 2140      & $<$ 0.05         & $<$ 0.13        & $<$ 0.09         \nl
C II] 2325      &     0.10 (0.03)  &     0.21 (0.07) &     0.16 (0.05)  \nl
[O II] 2470     & $<$ 0.03         & $<$ 0.05        & $<$ 0.04         \nl
He I 2945       & $<$ 0.02         & $<$ 0.03        & $<$ 0.02         \nl
He I 3188       & $<$ 0.02         & $<$ 0.03        & $<$ 0.02         \nl
[O II] 3727     &     2.65 (0.09)  &     3.14 (0.16) &     3.05 (0.14)  \nl
H 10 3795       & $<$ 0.06         & $<$ 0.07        & $<$ 0.07         \nl
H 9 3835        & $<$ 0.06         & $<$ 0.07        & $<$ 0.07         \nl
[Ne III] 3869   &     0.16 (0.03)  &     0.19 (0.04) &     0.18 (0.03)  \nl
H$\delta$ 4102  &     0.20 (0.03)  &     0.27 (0.04) &     0.27 (0.04)  \nl
H$\gamma$ 4340  &     0.40 (0.03)  &     0.47 (0.04) &     0.46 (0.04)  \nl
[O III] 4363    & $<$ 0.06         & $<$ 0.07        & $<$ 0.06         \nl
He I 4471       & $<$ 0.06         & $<$ 0.06        & $<$ 0.06         \nl
H$\beta$ 4861   &     1.00 (0.05)  &     1.00 (0.05) &     1.00 (0.05)  \nl
[O III] 4959    &     0.95 (0.06)  &     0.94 (0.06) &     0.94 (0.06)  \nl
[O III] 5007    &     2.93 (0.10)  &     2.87 (0.10) &     2.85 (0.10)  \nl
He I 5876       & $<$ 0.10         & $<$ 0.09        & $<$ 0.09         \nl
H$\alpha$ 6563  &     3.47 (0.11)  &     2.88 (0.14) &     2.85 (0.14)  \nl
[N II] 6584     &     0.52 (0.05)  &     0.44 (0.05) &     0.44 (0.05)  \nl
[S II] 6725     &     0.65 (0.08)  &     0.55 (0.07) &     0.54 (0.07)  \nl
                &                  &                 &                  \nl
A(H$\beta$) (mag.) &               & 0.58 (0.13)     &    0.80 (0.17)   \nl
                &                  &                 &                  \tablebreak
\cutinhead{NGC 5461}                 
[O III] 1666    & $<$ 0.12         & $<$ 0.54        & $<$ 0.22         \nl
N III] 1750     & $<$ 0.05         & $<$ 0.22        & $<$ 0.09         \nl
Si III] 1883    & $<$ 0.03         & $<$ 0.15        & $<$ 0.07         \nl
C III] 1909     &     0.06 (0.02)  &     0.30 (0.11) &     0.13 (0.04)  \nl
N II] 2140      & $<$ 0.02         & $<$ 0.19        & $<$ 0.08         \nl
C II] 2325      &     0.06 (0.02)  &     0.35 (0.12) &     0.18 (0.06)  \nl
[O II] 2470     &     0.08 (0.01)  &     0.30 (0.05) &     0.17 (0.02)  \nl
He I 2945       & $<$ 0.02         & $<$ 0.04        & $<$ 0.03         \nl
He I 3188       & $<$ 0.02         & $<$ 0.04        & $<$ 0.03         \nl
[O II] 3727     &     2.05 (0.06)  &     3.09 (0.12) &     2.86 (0.10)  \nl
H 10 3795       &     0.08 (0.02)  &     0.12 (0.03) &     0.11 (0.03)  \nl
H 9 3835        &     0.07 (0.02)  &     0.10 (0.03) &     0.10 (0.03)  \nl
[Ne III] 3869   &     0.12 (0.02)  &     0.17 (0.03) &     0.16 (0.03)  \nl
H$\delta$ 4102  &     0.22 (0.02)  &     0.30 (0.03) &     0.28 (0.03)  \nl
H$\gamma$ 4340  &     0.40 (0.02)  &     0.49 (0.03) &     0.48 (0.03)  \nl
[O III] 4363    & $<$ 0.03         & $<$ 0.04        & $<$ 0.04         \nl
He I 4471       & $<$ 0.03         & $<$ 0.04        & $<$ 0.03         \nl
H$\beta$ 4861   &     1.00 (0.03)  &     1.00 (0.03) &     1.00 (0.03)  \nl
[O III] 4959    &     0.98 (0.03)  &     0.95 (0.03) &     0.95 (0.03)  \nl
[O III] 5007    &     3.25 (0.10)  &     3.10 (0.10) &     3.12 (0.10)  \nl
He I 5876       &     0.12 (0.02)  &     0.09 (0.02) &     0.09 (0.02)  \nl
H$\alpha$ 6563  &     4.44 (0.14)  &     3.03 (0.12) &     2.96 (0.12)  \nl
[N II] 6584     &     0.63 (0.03)  &     0.43 (0.02) &     0.42 (0.02)  \nl
[S II] 6725     &     0.71 (0.04)  &     0.47 (0.03) &     0.46 (0.03)  \nl
                &                  &                 &                  \nl
A(H$\beta$) (mag.) &               & 1.39 (0.09)     &   1.91 (0.12)    \nl
                &                  &                 &                  \tablebreak
\cutinhead{NGC 5471}
[O III] 1666    &  $<$ 0.28        & $<$ 0.30        & $<$ 0.28         \nl
N III] 1750     &  $<$ 0.16        & $<$ 0.17        & $<$ 0.16         \nl
Si III] 1883    &      0.17 (0.08) &     0.18 (0.09) &     0.17 (0.08)  \nl
C III] 1909     &      0.47 (0.06) &     0.50 (0.08) &     0.48 (0.07)  \nl
N II] 2140      &  $<$ 0.06        & $<$ 0.07        & $<$ 0.06         \nl
C II] 2325      &  $<$ 0.07        & $<$ 0.08        & $<$ 0.07         \nl
[O II] 2470     &  $<$ 0.05        & $<$ 0.05        & $<$ 0.05         \nl
He I 2945       &  $<$ 0.04        & $<$ 0.04        & $<$ 0.04         \nl
He I 3188       &  $<$ 0.03        & $<$ 0.03        & $<$ 0.03         \nl
[O II] 3727     &      1.74 (0.06) &     1.77 (0.07) &     1.76 (0.07)  \nl
H 10 3795       &  $<$ 0.04        & $<$ 0.04        & $<$ 0.04         \nl
H 9 3835        &      0.05 (0.02) &     0.05 (0.02) &     0.05 (0.02)  \nl
[Ne III] 3869   &      0.50 (0.02) &     0.51 (0.02) &     0.50 (0.02)  \nl
H$\delta$ 4102  &      0.27 (0.02) &     0.27 (0.02) &     0.27 (0.02)  \nl
H$\gamma$ 4340  &      0.48 (0.02) &     0.48 (0.02) &     0.48 (0.02)  \nl
[O III] 4363    &      0.08 (0.02) &     0.08 (0.02) &     0.08 (0.02)  \nl
He I 4471       &  $<$ 0.04        & $<$ 0.04        & $<$ 0.04         \nl
H$\beta$ 4861   &      1.00 (0.04) &     1.00 (0.04) &     1.00 (0.04)  \nl
[O III] 4959    &      1.96 (0.06) &     1.96 (0.06) &     1.96 (0.06)  \nl
[O III] 5007    &      5.44 (0.16) &     5.43 (0.16) &     5.43 (0.16)  \nl
He I 5876       &      0.14 (0.03) &     0.14 (0.03) &     0.14 (0.03)  \nl
H$\alpha$ 6563  &      2.92 (0.10) &     2.88 (0.12) &     2.88 (0.12)  \nl
[N II] 6584     &      0.14 (0.04) &     0.14 (0.04) &     0.14 (0.04)  \nl
[S II] 6725     &  $<$ 0.13        & $<$ 0.13        & $<$ 0.13         \nl
                &                  &                 &                  \nl
A(H$\beta$) (mag.) &               & 0.05 (0.08)     &   0.06 (0.12)    \nl
                &                  &                 &                  \tablebreak
\cutinhead{VS 38}              
O III] 1666    & $<$ 0.04          & $<$ 0.1           & $<$ 0.06          \nl
N III] 1750    & $<$ 0.02          & $<$ 0.05          & $<$ 0.03          \nl
Si III] 1883   & $<$ 0.016         & $<$ 0.04          & $<$ 0.03          \nl
C III] 1909    & $<$ 0.017         & $<$ 0.05          & $<$ 0.03          \nl
N II] 2140     & $<$ 0.008         & $<$ 0.03          & $<$ 0.02          \nl
C II] 2325     &   0.035 (0.008)   &     0.10  (0.03)  &     0.07  (0.02)  \nl
[O II] 2470    & $<$ 0.012 (0.005) &     0.03  (0.01)  &     0.019 (0.008) \nl
He I 2945      & $<$ 0.009         & $<$ 0.02          & $<$ 0.01          \nl
He I 3188      & $<$ 0.028 (0.004) &     0.040 (0.006) &     0.035 (0.005) \nl
[O II] 3727    &   1.726 (0.053)   &     2.2   (0.1)   &     2.10  (0.09)  \nl
H 10 3795      &   0.033 (0.009)   &     0.04  (0.01)  &     0.04  (0.01)  \nl
H 9 3835       &   0.045 (0.009)   &     0.06  (0.01)  &     0.05  (0.01)  \nl
[Ne III] 3869  &   0.073 (0.009)   &     0.09  (0.01)  &     0.09  (0.01)  \nl
H$\delta$ 4102 &   0.246 (0.014)   &     0.29  (0.02)  &     0.28  (0.02)  \nl
H$\gamma$ 4340 &   0.448 (0.016)   &     0.51  (0.02)  &     0.50  (0.02)  \nl
[O III] 4363   & $<$ 0.018         & $<$ 0.02          & $<$ 0.02          \nl
He I 4471      &   0.044 (0.009)   &     0.05  (0.01)  &     0.05  (0.01)  \nl
H$\beta$ 4861  &   1.000 (0.032)   &     1.00  (0.03)  &     1.00  (0.03)  \nl
[O III] 4959   &   0.553 (0.021)   &     0.54  (0.02)  &     0.54  (0.02)  \nl
[O III] 5007   &   1.833 (0.057)   &     1.78  (0.06)  &     1.80  (0.06)  \nl
He I 5876      &   0.175 (0.012)   &     0.15  (0.01)  &     0.15  (0.01)  \nl  
[S III] 6312   & $<$ 0.033         & $<$ 0.03          & $<$ 0.03          \nl
H$\alpha$ 6563 &   3.927 (0.118)   &     3.1   (0.1)   &     3.1   (0.1)   \nl
[N II] 6584    &   0.494 (0.022)   &     0.39  (0.02)  &     0.39  (0.02)  \nl
[S II] 6725    &   0.202 (0.041)   &     0.16  (0.03)  &     0.16  (0.03)  \nl
               &                   &                   &                   \nl
A(H$\beta$)    &                   &  0.8   (0.1)      &     1.0   (0.2)   \nl
               &                   &                   &                   \tablebreak
\cutinhead{VS 44}              
O III] 1666    &   0.017 (0.014)  &     0.04  (0.03)  &     0.02  (0.02)   \nl
N III] 1750    & $<$ 0.016        & $<$ 0.04          & $<$ 0.03           \nl
Si III] 1883   & $<$ 0.015        & $<$ 0.04          & $<$ 0.03           \nl
C III] 1909    &   0.057 (0.007)  &     0.15  (0.03)  &     0.09  (0.01)   \nl
N II] 2140     & $<$ 0.007        & $<$ 0.03          & $<$ 0.02           \nl
C II] 2325     &   0.039 (0.004)  &     0.11  (0.02)  &     0.071 (0.009)  \nl
[O II] 2470    &   0.029 (0.004)  &     0.06  (0.01)  &     0.045 (0.007)  \nl
He I 2945      &   0.008 (0.002)  &     0.013 (0.003) &     0.010 (0.003)  \nl
He I 3188      &   0.022 (0.003)  &     0.031 (0.004) &     0.027 (0.004)  \nl
[O II] 3727    &   1.803 (0.054)  &     2.3   (0.1)   &     2.17  (0.08)   \nl
H 12 3750      &   0.015 (0.006)  &     0.019 (0.008) &     0.018 (0.007)  \nl
H 11 3770      &   0.017 (0.006)  &     0.021 (0.008) &     0.020 (0.007)  \nl
H 10 3795      &   0.024 (0.006)  &     0.030 (0.008) &     0.029 (0.007)  \nl
H 9 3835       &   0.042 (0.005)  &     0.053 (0.006) &     0.050 (0.006)  \nl
[Ne III] 3869  &   0.125 (0.006)  &     0.16  (0.01)  &     0.15  (0.01)   \nl
He I 4026      &   0.024 (0.005)  &     0.029 (0.006) &     0.028 (0.006)  \nl
H$\delta$ 4102 &   0.251 (0.008)  &     0.30  (0.01)  &     0.29  (0.01)   \nl
H$\gamma$ 4340 &   0.440 (0.013)  &     0.50  (0.02)  &     0.48  (0.02)   \nl
[O III] 4363   & $<$ 0.009        & $<$ 0.01          & $<$ 0.01           \nl
He I 4471      &   0.031 (0.005)  &     0.034 (0.006) &     0.033 (0.005)  \nl
H$\beta$ 4861  &   1.000 (0.030)  &     1.00  (0.03)  &     1.00  (0.03)   \nl
[O III] 4959   &   0.964 (0.030)  &     0.94  (0.03)  &     0.95  (0.03)   \nl
[O III] 5007   &   2.922 (0.088)  &     2.86  (0.09)  &     2.87  (0.09)   \nl
He I 5876      &   0.141 (0.007)  &     0.12  (0.01)  &     0.12  (0.01)   \nl
[S III] 6312   &   0.017 (0.005)  &     0.014 (0.004) &     0.014 (0.004)  \nl
H$\alpha$ 6563 &   3.787 (0.114)  &     3.0   (0.1)   &     3.1   (0.1)    \nl
[N II] 6584    &   0.459 (0.016)  &     0.37  (0.02)  &     0.37  (0.02)   \nl
He I 6678      &   0.053 (0.007)  &     0.042 (0.006) &     0.043 (0.006)  \nl
[S II] 6717    &   0.127 (0.008)  &     0.10  (0.01)  &     0.10  (0.01)   \nl
[S II] 6731    &   0.127 (0.008)  &     0.10  (0.01)  &     0.10  (0.01)   \nl
               &                  &                   &                    \nl
A(H$\beta$)    &                  &  0.7   (0.1)      & 0.9   (0.1)        \nl
               &                  &                   &                    \tablebreak
\cutinhead{VS 9 }             
O III] 1666    & $<$ 0.38         & $<$ 0.5         & $<$ 0.4           \nl
N III] 1750    & $<$ 0.27         & $<$ 0.4         & $<$ 0.3           \nl
Si III] 1883   & $<$ 0.17         & $<$ 0.2         & $<$ 0.2           \nl
C III] 1909    &   0.363 (0.082)  &     0.4  (0.2)  &     0.3  (0.1)    \nl
N II] 2140     & $<$ 0.10         & $<$ 0.2         & $<$ 0.1           \nl
C II] 2325     &   0.102 (0.052)  &     0.14 (0.08) &     0.12 (0.06)   \nl
[O II] 2470    & $<$ 0.10         & $<$ 0.1         & $<$ 0.1           \nl
He I 2945      & $<$ 0.06         & $<$ 0.07        & $<$ 0.07          \nl
He I 3188      & $<$ 0.06         & $<$ 0.07        & $<$ 0.07          \nl
[O II] 3727    &   2.004 (0.085)  &     2.2  (0.2)  &     2.1  (0.1)    \nl
H 10 3795      & $<$ 0.15         & $<$ 0.2         & $<$ 0.2           \nl
H 9 3835       & $<$ 0.15         & $<$ 0.2         & $<$ 0.2           \nl
[Ne III] 3869  &   0.288 (0.062)  &     0.31 (0.07) &     0.30 (0.07)   \nl
H$\delta$ 4102 & $<$ 0.26         & $<$ 0.3         & $<$ 0.3           \nl
H$\gamma$ 4340 &   0.465 (0.064)  &     0.48 (0.07) &     0.48 (0.07)   \nl
[O III] 4363   & $<$ 0.12         & $<$ 0.1         & $<$ 0.1           \nl
He I 4471      & $<$ 0.13         & $<$ 0.1         & $<$ 0.1           \nl
H$\beta$ 4861  &   1.000 (0.068)  &     1.00 (0.07) &     1.00 (0.07)   \nl
[O III] 4959   &   1.225 (0.069)  &     1.22 (0.07) &     1.22 (0.07)   \nl
[O III] 5007   &   4.092 (0.137)  &     4.0  (0.1)  &     4.0  (0.1)    \nl
He I 5876      & $<$ 0.10         & $<$ 0.1         & $<$ 0.1           \nl
[S III] 6312   & $<$ 0.11         & $<$ 0.1         & $<$ 0.1           \nl
H$\alpha$ 6563 &   3.125 (0.111)  &     2.9  (0.2)  &     2.9  (0.2)    \nl
[N II] 6584    &   0.393 (0.062)  &     0.37 (0.06) &     0.37 (0.06)   \nl
[S II] 6725    & $<$ 0.22         & $<$ 0.2         & $<$ 0.2           \nl
               &                  &                 &                   \nl
A(H$\beta$)    &                  &    0.2  (0.2)   &     0.3  (0.3)    \nl
\enddata
\end{deluxetable}

\clearpage

\begin{deluxetable}{lccc}
\tablewidth{35pc}
\tablecaption{Adopted Electron Densities and Electron Temperatures }
\tablehead{
\colhead{Object} & \colhead{n$_e$ (cm$^{-3}$)} & \colhead{T[O~III] (K)} & 
\colhead{T[O~II] (K)} \nl
}
\startdata
NGC 5455 & 100 &  9700$\pm$500 &  9800$\pm$600 \nl
NGC 5461 & 100 &  9000$\pm$300 &  9300$\pm$500 \nl
NGC 5471 & 100 & 13200$\pm$300 & 12200$\pm$500 \nl
VS 38    & 100 &  7600$\pm$600 &  8300$\pm$600 \nl
VS 44    & 600 &  8700$\pm$400 &  9100$\pm$600 \nl
VS  9    & 100 & 11700$\pm$400 & 11200$\pm$400 \nl
\enddata
\end{deluxetable}

\clearpage

\begin{deluxetable}{lccccccc}
\tablewidth{43pc}
\tablecaption{Ionic Abundances from FOS Observations }
\tablehead{
\colhead{Object} & \colhead{log {$O^+ \over H^+$}} & \colhead{log {$O^{+2} \over H^+$}} & \colhead{log $C^+\over H^+$} & \colhead{log $C^+\over H^+$} &  
\colhead{log $C^{+2}\over H^+$} & \colhead{log $C^{+2}\over H^+$} \\  
\colhead{} &\colhead{} & \colhead{} & \colhead{$R_V$ = 3.1} & \colhead{$R_V$ = 5} & \colhead{$R_V$ = 3.1} & \colhead{$R_V$ = 5} \nl
}
\tablecolumns{7}
\startdata
NGC 5455 & $-$3.89$\pm$0.07 & $-$3.93$\pm$0.05 & $-$4.32$\pm$0.14 & $-$4.44$\pm$0.14 & $-$4.05$\pm$0.16 & $-$4.18$\pm$0.15 & \nl
NGC 5461 & $-$3.79$\pm$0.05 & $-$3.78$\pm$0.03 & $-$3.93$\pm$0.14 & $-$4.22$\pm$0.14 & $-$3.73$\pm$0.15 & $-$4.10$\pm$0.13 & \nl
NGC 5471 & $-$4.55$\pm$0.04 & $-$4.06$\pm$0.03 & $<$ $-$5.3 & $<$ $-$5.4 & $-$4.70$\pm$0.07 & $-$4.72$\pm$0.07 & \nl
VS 38    & $-$3.69$\pm$0.07 & $-$3.76$\pm$0.07 & $-$4.12$\pm$0.15 & $-$4.30$\pm$0.15 & $-$4.00$\pm$0.16 & $-$4.16$\pm$0.16 & \nl
VS 44    & $-$3.83$\pm$0.06 & $-$3.78$\pm$0.04 & $-$4.38$\pm$0.11 & $-$4.57$\pm$0.10 & $-$3.92$\pm$0.11 & $-$4.15$\pm$0.09 & \nl
VS  9    & $-$4.28$\pm$0.05 & $-$4.04$\pm$0.04 & $-$4.85$\pm$0.20 & $-$4.92$\pm$0.18 & $-$4.57$\pm$0.18 & $-$4.66$\pm$0.13 & \nl
\enddata
\end{deluxetable}

\clearpage

\begin{deluxetable}{lcccc}
\tablewidth{34pc}
\tablecaption{Total Abundances for NGC 2403 and M101 H II Regions}
\tablehead{
\colhead{Object} & \colhead{log O/H} & \colhead{log C/O} & \colhead{log C/O} & \colhead{log N/O} \\ 
\colhead{} & \colhead{} & \colhead{$R_V$ = 3.1} & \colhead{$R_V$ = 5} & \colhead{} \nl
}
\startdata
NGC 5455 & $-$3.61$\pm$0.03 & $-$0.26$\pm$0.14 & $-$0.38$\pm$0.13 & $-$1.18$\pm$0.10 \nl
NGC 5461 & $-$3.49$\pm$0.03 & $-$0.03$\pm$0.13 & $-$0.37$\pm$0.12 & $-$1.13$\pm$0.06 \nl
NGC 5471 & $-$3.94$\pm$0.03 & $-$0.67$\pm$0.05 & $-$0.70$\pm$0.05 & $-$1.33$\pm$0.06 \nl
VS 38    & $-$3.42$\pm$0.05 & $-$0.33$\pm$0.12 & $-$0.44$\pm$0.11 & $-$1.08$\pm$0.03 \nl
VS 44    & $-$3.50$\pm$0.04 & $-$0.29$\pm$0.08 & $-$0.51$\pm$0.07 & $-$1.19$\pm$0.03 \nl
VS  9    & $-$3.84$\pm$0.03 & $-$0.55$\pm$0.19 & $-$0.63$\pm$0.15 & $-$1.21$\pm$0.03 \nl
\enddata
\end{deluxetable}

%

\end{document}